\documentstyle[preprint,prd,aps]{revtex}
\begin{document}
%\draft
%%%%%End of Preamble
%%%%Start of Text%%%%%%%%%%%%%%%%%%%%%%%%%%%%%%%%%%%%%%%%%%%%%%%%%%%%%%%
\preprint{
\vbox{\halign{&##\hfil\cr
		& OHSTPY-HEP-T-96-018	\cr
		& October 1996	\cr
		& \ 	\cr
		& \ 	\cr
}}}
		
\title{Dimensional Regularization in Quarkonium Calculations}

\author{Eric Braaten and Yu-Qi Chen}
\address{Physics Department, Ohio State University, Columbus OH 43210}

\maketitle

\begin{abstract}
Dimensional regularization is incompatible with the standard covariant 
projection methods that are used to calculate the short-distance 
coefficients in inclusive heavy quarkonium production and annihilation 
rates.  A new method is developed that allows dimensional regularization 
to be used consistently to regularize the infrared and ultraviolet 
divergences that arise in these perturbative calculations.  
We illustrate the method by calculating the leading color-octet terms 
and the leading color-singlet terms in the gluon fragmentation 
functions for arbitrary quarkonium states.  We resolve a
discrepancy between two  previous calculations of the 
gluon fragmentation functions for the spin-triplet P-wave quarkonium states.     
\end{abstract}


\vfill \eject

\narrowtext

\section{Introduction}

A rigorous theoretical framework for calculating 
inclusive production cross sections (and inclusive annihilation decay rates)
for heavy quarkonium in QCD has recently been developed \cite{B-B-L}. 
The cross section (or decay rate) is expressed as a sum of the product 
of short-distance coefficients and long-distance matrix elements. 
The short-distance coefficients can be calculated by 
perturbation theory to any order in $\alpha_s(m_c)$, 
the running coupling constant  at the energy scale of the heavy 
quark mass. The long-distance factors are expressed as matrix elements 
in nonrelativistic QCD (NRQCD), an effective field theory
that reproduces QCD to any desired order in the velocity of the 
heavy quark.  The matrix elements scale in a definite way with $v$,
the typical relative velocity of the heavy quark in quarkonium.
The NRQCD factorization formalism therefore allows the  
inclusive production cross section to be calculated systematically as a 
double expansion in $\alpha_s(m_c)$ and $v$. 

In calculating the short-distance coefficients beyond 
leading order in $\alpha_s$,
ultraviolet divergences and infrared divergences inevitably 
arise and need to be regulated.  In perturbative calculations,
the most convenient method for regulating 
both ultraviolet and infrared divergences 
is dimensional regularization.  The main advantage of this method
is that it preserves many of the symmetries of a field theory,
including gauge invariance and Lorentz invariance. 
It also often leads to simpler expressions for the finite parts 
of a perturbative calculation.
One feature of dimensional regularization that is particularly
convenient in a nonrenormalizable effective field theory like NRQCD
is that it automatically sets power ultraviolet divergences to zero. 
This makes it unnecessary to explicitly subtract these divergences 
as part of the renormalization procedure.
Dimensional regularization has been
widely used in perturbative calculations of the cross sections 
and decay rates of elementary particles.
It has also been used in some perturbative calculations involving
quarkonium, but the consistency of the method has not been carefully examined.

Most calculations of production cross sections and annihilation decay rates 
for heavy quarkonium have been carried out using the
covariant projection method \cite{K-K-S}.  This method involves
the projection of a $c \bar c$ pair onto states with definite 
total angular momentum $J$.  For orbital angular momentum 
$L=1$ and higher, the projections are specific to 3 dimensions.  
In dimensional regularization, momentum integrals are analytically 
continued to $N = 3 - 2 \epsilon$ dimensions.  There is therefore a 
potential inconsistency in combining the covariant projection method 
with dimensional regularization.  Moreover, 
the projections for spin-singlet states, such as $^1S_0$ and $^1P_1$, 
involve the $\gamma_5$ matrix, whose extension to $N$ dimensions is 
problematic.

We have recently developed an alternative method for calculating
production cross sections and annihilation decay rates 
for heavy quarkonium which fully exploits the NRQCD factorization 
framework \cite{Braaten-Chen}.  We refer to it as the 
{\it threshold expansion method}.  In this method, 
 a quantity that is closely related to 
the cross section for the production of a 
$c\bar{c}$ pair with total momentum $P$ is calculated using 
perturbation theory in full QCD and expanded in powers of the 
relative 3-momentum ${\bf q}$ of the $ c \bar c$ pair.
Matrix elements of certain 4-fermion operators in  NRQCD
are also calculated using perturbation theory and expanded 
 around the threshold ${\bf q}= 0$. 
The short-distance coefficients in the factorization formula
are then determined by matching  these perturbative expressions.
Finally, the NRQCD matrix elements for quarkonium states are
simplified by using rotational symmetry, heavy-quark 
spin symmetry, and the vacuum saturation approximation.

In this paper, we generalize this method to $N$ spatial dimensions,
so that dimensional regularization can be consistently used to 
regularize infrared and ultraviolet divergences.
The perturbative calculation of the QCD 
 side of the matching condition and its expansion around the threshold 
generalize easily to $N$ dimensions.  The perturbative calculation and 
threshold expansion of the NRQCD matrix elements can also be carried out 
in $N$ dimensions, provided that the matrix elements are defined 
in a way that generalizes to $N$ dimensions.
By matching the cross section and the matrix elements, we obtain 
the short-distance coefficients.  After renormalization of 
coupling constants in QCD and NRQCD, the short-distance coefficients 
may have poles in $N-3$, which must be removed
by renormalization of the 4-fermion operators in NRQCD.
One must take care to avoid simplifying the matrix elements of these 
operators using identities that are specific to 3 dimensions until
after these renormalizations have been carried out.

In Section II, we review the threshold expansion method
of Ref. \cite{Braaten-Chen}, generalizing it to $N$ spatial dimensions.
In Section III, we give a definition of the gluon fragmentation 
function for quarkonium states that is particularly convenient 
for low order calculations.
We then present a matching prescription for calculating the short-distance 
coefficients in the factorization formula for these fragmentation functions. 
In Sections IV and V, we illustrate our calculational method by
calculating the color-octet terms in the 
gluon fragmentation function 
 for a general quarkonium state 
to order $\alpha_s$ and 
the color-singlet terms to order $\alpha_s^2$.
The color-singlet calculation involves a pole in $N-3$ which must be removed
by the renormalization of an NRQCD matrix element.
We apply our general formula for the gluon fragmentation function to 
S-wave 
 and P-wave states in Section VI. 
We resolve a discrepancy between two previous calculations of the 
gluon fragmentation function for $^3P_J$ states.

\section{Threshold Expansion Method}

We begin by reviewing the threshold expansion method of 
Ref. \cite{Braaten-Chen}, generalizing it to $N$ dimensions.
Cartesian vectors and Lorentz vectors are 
extended to $N$ and $N+1$ components, respectively.  The inclusive
cross section for producing the heavy quarkonium $H$ with momentum $P$
can be written in a factorized form \cite{B-B-L}:
\begin{equation}
\sum_X d \sigma(12 \to H(P) + X)
\;=\; {1 \over 4 E_1 E_2 v_{12}} \; {d^N P \over (2 \pi)^N 2 E_P} \;
	\sum_{mn} C_{mn} \langle {\cal O}_{mn}^H \rangle,
\label{fact-sig}
\end{equation}
where $E_P = \sqrt{M_H^2 + {\bf P}^2}$.
The coefficients $C_{mn}$ take into account the effects of short 
distances of order $1/m_c$ or smaller, 
and therefore can be calculated as perturbation series in the QCD coupling 
constant $\alpha_s(m_c)$.  The matrix elements 
$\langle {\cal O}_{mn}^H \rangle$ are expectation values in the 
NRQCD vacuum of local 4-fermion operators that have the structure
\begin{equation}
{\cal O}_{mn}^H
\;=\; \chi^\dagger {{\cal K}'}_m^\dagger \psi \; 
	{\cal P}_H \;
	\psi^\dagger {\cal K}_n \chi ,	
\label{O-NRQCD}
\end{equation}
where $\psi$ and $\chi$ are the field operators for the heavy quark and 
antiquark in NRQCD, and ${\cal K}_n$ and ${{\cal K}'}_m^\dagger$
are products of a color matrix ($1$ or $T^a$), 
a spin matrix, and a polynomial in the gauge covariant 
derivative ${\bf D}$ in $N$ dimensions.  
The spin matrix is either the unit matrix or a polynomial in
the Pauli matrices $\sigma^i$.
The Pauli matrices in $N$ dimensions satisfy the anticommutation
relations
\begin{eqnarray}
\left \{\;
	\sigma_i,\; \sigma_j 
\;\right \} 
 & = & 2 \delta_{ij}, ~~~~~~~~~~i,j=1,\ldots, N.
\label{anti-commu} 
\end{eqnarray} 
In 3 dimensions, they also satisfy the commutation relations
\begin{eqnarray}
\left [\;
     \sigma_i,\; \sigma_j 
\;\right ] 
 & = & 2 i\, \epsilon_{ijk}\; \sigma_k , ~~~~~~~~~~ i,j,k=1,2,3.
\label{commu}
\end{eqnarray}
Using both the relations (\ref{anti-commu}) and (\ref{commu}),
the spin matrices in  ${\cal K}_n$ and ${{\cal K}'}_m^\dagger$
in (\ref{O-NRQCD}) can be reduced to a linear combination of 
1 and $\sigma^i$. 
However, the commutation relations (\ref{commu}) can be used to simplify 
the NRQCD matrix elements only after all poles in $N-3$
have been removed from the short-distance coefficients.
Until these poles are removed, we must allow for additional
spin matrices in ${\cal K}_n$ and ${{\cal K}'}_m^\dagger$, 
such as $[ \sigma_i, \sigma_j ]$ and
$\{ [ \sigma_i, \sigma_j ], \sigma_k \}$.
The projection operator
${\cal P}_H$ in (\ref{fact-sig}) can be written
\begin{equation}
{\cal P}_H 
\;=\; \sum_S \Big | H({\bf P} = 0) + S \Big \rangle 
	\Big \langle H({\bf P} = 0) + S \Big | ,
\label{P_H}
\end{equation}
where the sum is
over soft hadron states $S$ whose total energy is less than the 
ultraviolet cutoff $\Lambda$ of NRQCD.  This operator 
projects onto the subspace of states which in the asymptotic feature
include the quarkonium state $H$ at rest plus soft hadrons.
The standard relativistic normalization of the states in (\ref{P_H}) is
\begin{equation}
\Big \langle H({\bf P}') \Big | H({\bf P}) \Big\rangle 
\;=\; 2 E_P \; (2 \pi)^N  \delta^N ({\bf P} - {\bf P}') .	
\label{norm_H}
\end{equation}
With this normalization of states, the projection operator 
${\cal P}_H$ has energy dimension $1-N$. 

The short distance coefficients $C_{mn}$ in (\ref{fact-sig})
can be determined by matching perturbative calculations of the 
corresponding process in which the quarkonium $H$ is replaced
by a $c \bar{c}$ pair. 
Let $c \bar c(P,{\bf q},\xi,\eta)$ represent a state that consists of a $c$ 
and a $\bar c$ with total momentum $P$,  spatial momenta $\pm {\bf q}$ in the 
$c \bar c$ rest frame, and spin and color states specified by the spinors
$\xi$ and $\eta$.  Color and spin indices on these
spinors are suppressed.  The number of spin components for these
Pauli spinors is that appropriate to  Pauli spinors  in $N$ spatial 
dimensions. Using the abbreviated notation 
$c \bar c(P) \equiv c \bar c(P,{\bf q},\xi,\eta)$ and
$c \bar c'(P) \equiv c \bar c(P,{\bf q}',\xi',\eta')$, the matching condition
 in the threshold expansion method of Ref. \cite{Braaten-Chen} is 
\begin{eqnarray}
\sum_X (2 \pi)^{N+1} \delta^{N+1}(k_1 + k_2 - P - k_X) \;
	\left( {\cal T}_{ c \bar c'(P) + X} \right)^* \; 
	{\cal T}_{ c \bar c(P) + X} \Big|_{pQCD}
\nonumber \\
\;=\; \sum_{mn} C_{mn} \;
	\langle \chi^\dagger {{\cal K}'}_m^\dagger \psi \; 
	{\cal P}_{c \bar c',c \bar c} \;
	\psi^\dagger {\cal K}_n \chi \rangle \Big|_{pNRQCD},
\label{match-sig}
\end{eqnarray}
where $k_1$ and $k_2$ are the momenta of the incoming particles 
and $k_X$ is the sum of the momenta 
of all the outgoing particles except the $c$ and $\bar c$.
The operator ${\cal P}_{c \bar c',c \bar c}$ in the matrix 
element in (\ref{match-sig}) is defined by 
\begin{equation}
{\cal P}_{c \bar c',c \bar c} 
\;=\; \sum_S \Big | c({\bf q}',\xi') \bar c(-{\bf q}',\eta') + S \Big \rangle 
	\Big \langle c({\bf q},\xi) \bar c(-{\bf q},\eta) + S \Big | .	
\label{P_cc}
\end{equation}
The sum is over soft parton states whose total energy is  less than the 
ultraviolet cutoff $\Lambda$ of NRQCD. 
The standard relativistic normalization is
\begin{equation}
\Big \langle c({\bf q}_1',\xi') \bar c({\bf q}_2',\eta')
	 \Big | c({\bf q_1},\xi) \bar c({\bf q_2},\eta)  \Big \rangle 
\;=\; 4 E_{q_1} E_{q_2} \; (2 \pi)^{2 N}  \delta^N ({\bf q}_1 - {\bf q}_1')
	\delta^N({\bf q}_2 - {\bf q}_2') \; 
	\xi^\dagger \xi' {\eta'}^\dagger \eta ,
\label{cc-norm}
\end{equation}
where $E_q = \sqrt{m_c^2 + {\bf q}^2}$.  The spinors are normalized so that 
 $\xi^\dagger \xi = 1$, and similarly for $\eta$, 
$\xi'$ and $\eta'$.
In expressions like $\xi^\dagger \xi'$,
both the spin and color indices are contracted.
With the  normalization (\ref{cc-norm}), the projection operator
${\cal P}_{c \bar c',c \bar c}$ has dimension $2-2N$.
To carry out the matching procedure, the left side of (\ref{match-sig}) is
calculated using perturbation theory in full QCD, and then expanded 
in powers of ${\bf q}$ and ${\bf q}'$. The matrix elements on the right 
side of (\ref{match-sig}) are 
calculated using perturbation theory in NRQCD, and then expanded 
in powers of ${\bf q}$ and ${\bf q}'$.  The short-distance coefficients
$C_{mn}$ are obtained by matching the terms in the expansions
in ${\bf q}$ and ${\bf q}'$ order by order in $\alpha_s$.

In the perturbative calculations of the matching condition 
(\ref{match-sig}), infrared and ultraviolet divergences can appear 
on both sides of the equation.  Since NRQCD is constructed to be 
equivalent to full QCD at low energies, the infrared divegences 
on both sides must match.  They therefore cancel in the 
short-distance coefficients $C_{mn}$.  Any ultraviolet divergences
on the left side are eliminated by renormalization 
of the QCD coupling constant and the heavy quark mass.
On the right side, some of the ultraviolet divergences are eliminated
by renormalization of the gauge coupling constant and other parameters 
in the NRQCD lagrangian.  The remaining ultraviolet divergences are 
removed by renormalization of the 4-fermion operators of NRQCD.
 
The matching calculations are particularly simple if dimensional 
regularization is used to regulate both infrared and ultraviolet divergences.
With dimensional regularization, power infrared diverences 
and power ultraviolet divergences are automatically set equal to zero.
The only divergences which remain are logarithmic divergences, which
appear as poles in $\epsilon = (3-N)/2$. 
With dimensional regularization, the NRQCD side of the matching condition 
is especially simple, because radiative corrections to the matrix elements 
vanish identically.  The reason for this is that one must expand
the integrand of the radiative correction using a nonrelativistic 
expansion in the loop momentum before integrating.  Since the integrand
is also expanded in powers of ${\bf q}$ and ${\bf q}'$, there is no
momentum scale in the dimensionally regularized integral 
and it therefore vanishes.  The radiative corrections have both 
infrared and ultraviolet divergences,
but the infrared poles in $\epsilon$ cancel the ultraviolet poles.
Thus the only contributions to the NRQCD side of the matching condition
are the tree-level contributions of the matrix elements, including those
matrix elements that arise from counterterms 
associated with operator renormalization.  The poles in $\epsilon$
from the coefficients of the counterterm matrix elements will match 
the infrared poles in $\epsilon$ on the QCD side of the matching condition.
The short-distance coefficients of the renormalized matrix elements
 in the factorization formula 
depend on the operator renormalization scheme
 for NRQCD. 
The renormalized matrix elements in the minimal subtraction scheme 
are defined by the condition that the coefficients of the 
counterterms in the expression for 
 the bare operators in terms of renormalized operators 
are pure poles in $\epsilon$.
In the more conventional $\overline{MS}$ renormalization scheme, 
the coefficients of the counterterms are multiples of  
$1/\epsilon + \log(4 \pi) - \gamma$,
where $\gamma$ is Euler's constant.
One must be wary of simplifying the matrix elements using identities 
that are specific to 3 dimensions until after these renormalizations 
have been carried out.

\section{Factorization of Fragmentation Functions}
 
To illustrate the use of dimensional regularization in 
the threshold expansion method, we will calculate the terms of
lowest order in $\alpha_s$ in the gluon fragmentation function 
for an arbitrary quarkonium state. 
The fragmentation function $D_{i \to H}(z)$ gives the probability
for a jet initiated by a high energy parton $i$ to include the 
hadron $H$ carrying a fraction $z$ of the jet momentum.
The factorization formula (\ref{fact-sig}), when applied to 
the fragmentation function for a gluon into a heavy quarkonium state $H$, 
yields the factorized form
\begin{equation}
D_{g \to H}(z)
\;=\; \sum_{mn} d_{mn}(z) \langle {\cal O}_{mn}^H \rangle ,
\label{fact-frag}
\end{equation}
where the short-distance coefficients $d_{mn}(z)$ can be calculated 
using perturbation theory.

Fragmentation functions can be calculated directly
from the gauge-invariant field-theoretic definitions \cite{Ma-def}.
This method is particularly advantageous for higher order calculations.
We choose instead to follow as closely as possible
the calculational strategy introduced in Ref. \cite{Braaten-Yuan:Swave},
in which the fragmentation function is obtained by calculating 
the decay rate of a virtual parton in the infinite momentum frame
and in an appropriate axial gauge. 
We consider the fragmentation of a gluon into a quarkonium state $H$.
We take the gluon momentum to be $\ell = (\ell_0,0,\ldots,0,\ell_N)$.
The infinite momentum frame is defined by the limit 
$\ell_0,\ell_N \to \infty$ with 
$\ell^2= \ell_0^2 - \ell_N^2$ of order $m_c^2$.
We choose an axial gauge with reference vector $n = (1,0,\ldots,0,-1)$,
so that the gluon propagator is 
\begin{equation}
G_{\mu\nu}(\ell) \;=\; {1 \over \ell^2 + i \epsilon} 
           \left( - g_{\mu\nu} 
          + {n_\mu \ell_\nu + \ell_\mu n_\nu \over n \cdot \ell} \right).
\label{axial}
\end{equation}
We also introduce ${\bar n} = (1,0,\ldots,0,+1)$.
Let ${\cal A}^{\mu a}_{g^* \to H(P)+X}$ be the amplitude for 
a virtual gluon to decay into a quarkonium state $H$
plus additional final state particles $X$ whose total momentum is $k_X$.
The gluon fragmentation function can be defined by
\begin{eqnarray}
D_{g \to H}(z) &=& 
\int {d \ell^2 \over 2 \pi} \int {d^N P \over (2 \pi)^N 2 E_P} \sum_X \;
	(2 \pi)^{N+1} \delta^{N+1}(\ell - P - k_X) 
	\; \delta \left( z - {P \cdot n \over \ell \cdot n} \right)
\nonumber \\
&& \times {1 \over 8 (N-1)}{1 \over (\ell^2)^2}
	\left( {\cal A}^{\nu a}_{g^* \to H(P)+X} \right)^* 
	{\cal A}^{\mu a}_{g^* \to H(P)+X}
	\left( - g_{\mu\nu} 
	+ {n_\mu {\bar n}_\nu + {\bar n}_\mu n_\nu \over 2} \right).
\label{frag}
\end{eqnarray}
The sum over additional final state particles $X$ includes integration
over their phase space.  The last factor in (\ref{frag})
is the numerator of the gluon propagator in (\ref{axial})
evaluated at the positive energy  pole 
$\ell_0 = \ell_N + i \epsilon$.  It can also be written in the form

\begin{equation}
- g_{\mu\nu} 
	+ {n_\mu {\bar n}_\nu + {\bar n}_\mu n_\nu \over 2} 
\;=\; \sum_{i=1}^{N-1}  \epsilon_\mu^{(i)*}\epsilon_\nu^{(i)} ,
\label{gluon-sum}
\end{equation}
where $\epsilon_\mu^{(i)}$, $i=1,\ldots,N-1$, are the
transverse polarization 
vectors for a real gluon whose momentum is proportional to $\bar n$.
The factor of $1/(8(N-1))$ in (\ref{frag}) comes from averaging
over the color and polarization states of the decaying gluon.
In calculations beyond leading order, there are infrared
poles in $N-3$ that arise from the splitting of the gluon into 
collinear partons.  They should be absorbed into the Altarelli-Parisi evolution
of the fragmentation functions.  This complication will not enter 
into the leading-order calculations presented in this paper. 

The short-distance coefficients $d_{mn}(z)$ defined by the 
factorization formula (\ref{fact-frag}) can be obtained by a 
matching prescription analogous to (\ref{match-sig})
for the cross section:
\begin{eqnarray}
\int {d \ell^2 \over 2 \pi} \int {d^N P \over (2 \pi)^N 2 E_P} \sum_X \;
	(2 \pi)^{N+1} \delta^{N+1}(\ell - P - k_X) \;
	\delta \left( z - {P \cdot n \over \ell \cdot n} \right) 
\hspace{1in}
\nonumber \\
\times {1 \over 8 (N-1)}{1 \over (\ell^2)^2}
	\left( - g_{\mu\nu} 
	+ {n_\mu {\bar n}_\nu + {\bar n}_\mu n_\nu \over 2} \right)
	\left( {\cal A}^{\nu a}_{g^* \to c \bar c'(P) + X} \right)^* 
	{\cal A}^{\mu a}_{g^* \to c \bar c(P) + X} \Big|_{pQCD}
\nonumber \\
\;=\; \sum_{mn} d_{mn}(z) \;
	\langle \chi^\dagger {{\cal K}'}_m^\dagger \psi \; 
	{\cal P}_{c \bar c',c \bar c} \;
	\psi^\dagger {\cal K}_n \chi \rangle \Big|_{pNRQCD} \,.
\label{match-frag}
\end{eqnarray}
The left side is calculated using perturbative QCD and 
expanded around the thresholds ${\bf q} = {\bf q}' = 0$.
The matrix elements on the right side are calculated using 
perturbative NRQCD and also
expanded around the thresholds.  Matching the expansions in 
 ${\bf q}$ and ${\bf q}'$ order by order in $\alpha_s$, we can 
determine the short-distance coefficients $d_{mn}(z)$.

\section{Color-octet Terms at order $\alpha_s$}

The leading color-octet terms in the 
gluon fragmentation function $D_{g \to H}(z)$
have short-distance coefficients of order $\alpha_s$.  
These terms come from the decay of the virtual gluon through
the process $g^* \to c \bar c$.  For this simple case, there are
no additional final state particles $X$ 
in the left side of the 
matching condition (\ref{match-frag}).  The energy-momentum-conserving
delta function forces $P = \ell$ and 
the QCD side of 
the matching condition collapses to
\begin{equation}
\delta(1-z) {1 \over 8 (N-1) (P^2)^2}
	\left( - g_{\mu\nu} 
	+ {n_\mu {\bar n}_\nu + {\bar n}_\mu n_\nu \over 2} \right)
	\left( {\cal A}^{\nu a}_{g^* \to c \bar c'(P)} \right)^* 
	{\cal A}^{\mu a}_{g^* \to c \bar c(P)}
	\Big|_{pQCD}\;.
\label{match-ccbar}
\end{equation}
At leading order in $\alpha_s$, the amplitude for 
$g^*(\ell) \to c(p) \bar c(\bar p)$ is
 given by the Feynman diagram in Fig. 1: 
\begin{equation}
{\cal A}^{\mu a}_{g^* \to c \bar c(P)}
\;=\; g_s \mu^\epsilon \; \bar u(p) \gamma^\mu T^a v({\bar p}) .
\label{M-ccbar}
\end{equation}
The coupling constant in (\ref{M-ccbar}) has been written
$g_s \mu^\epsilon$, where $\epsilon = (3-N)/2$ and $\mu$ is the 
scale of dimensional regularization, so that $g_s$ remains dimensionless
in $N$ dimensions. 
The momenta of the $c$ and $\bar c$ are $p = {1 \over 2}P + L {\bf q}$ and 
$\bar p = {1 \over 2}P - L {\bf q}$, where
$L^\mu_i$ is the matrix that boosts a spacelike vector in the rest
frame of the $c \bar c$ pair to the frame in which the pair has 
total momentum $P$.  The components of the boost matrix are given in 
(\ref{L-boost}) of Appendix A.  
The spinor factor in (\ref{M-ccbar}) is expressed in terms of 
nonrelativistic Pauli spinors  in (\ref{bispinor-gam}).  
The amplitude becomes
\begin{equation}
{\cal A}^{\mu a}_{g^* \to c \bar c(P)}
\;=\; 2 m_c g_s \mu^\epsilon \; L^\mu_i \; \xi^\dagger \sigma^i T^a \eta \,.
\label{M-ccbar:NR}
\end{equation}
Inserting this into the 
QCD side of the matching condition (\ref{match-ccbar}), it 
reduces to  
\begin{equation}
\delta(1-z) {\pi \alpha_s \mu^{2\epsilon} \over 8 (N-1) m_c^2}  
\left( - g_{\mu\nu} 
	+ {n_\mu {\bar n}_\nu + {\bar n}_\mu n_\nu \over 2} \right) \;
 L^\nu_j \; L^\mu_i \;
\eta'^\dagger \sigma^j T^a \xi' \xi^\dagger \sigma^i T^a \eta \,.
\label{match-ccbar:1}
\end{equation}
Using the explicit expression for the boost matrices in (\ref{L-boost}),
the contraction of $L^\nu_j L^\mu_i$ with $n_\nu \bar n_\mu$ reduces to
$(n \cdot L)_j (\bar n \cdot L)_i  = - {\hat z}^j {\hat z}^i$,
where $\hat{\bf z}$ is the unit vector in the $N$'th coordinate direction.
Using also the identity (\ref{id:gL}), the 
QCD side of the 
matching condition reduces to 
\begin{equation}
\delta(1-z) {\pi \alpha_s \mu^{2 \epsilon} \over 8 (N-1) m_c^2} \; 
\left( \delta^{ij} - {\hat z}^i {\hat z}^j \right)
\eta'^\dagger \sigma^j T^a \xi' \xi^\dagger \sigma^i T^a \eta .
\label{match-ccbar:2}
\end{equation}

We now consider the NRQCD side of the matching condition (\ref{match-frag}). 
The spinor factor in (\ref{match-ccbar:2})
can be identified as the 
 expansion to leading order in $\alpha_s$ and to linear order in 
${\bf q}$ and ${\bf q}'$ of the following 
NRQCD matrix element:
\begin{equation}
\langle \chi^\dagger \sigma^j T^a \psi \, 
	{\cal P}_{c \bar c',c \bar c} \,
        \psi^\dagger \sigma^i T^a \chi \rangle \Big|_{pNRQCD} 
\;\approx\; 4 m_c^2 \; 
	\eta'^\dagger \sigma^j T^a \xi' \xi^\dagger \sigma^i T^a \eta .
\label{matel-ij}
\end{equation}
The tree-level expression for the matrix element in (\ref{matel-ij})
is represented diagramatically in Fig. 2.
The dot represents the operators $\psi^\dagger \sigma^i T^a \chi$
and $\chi^\dagger \sigma^j T^a \psi$, which create and annihilate
$c \bar c$ pairs from the vacuum.  The 2 lines emerging from the right side
of the diagram represent the $c$ and $\bar c$ in the bra of the
projection operator ${\cal P}_{c \bar c',c \bar c}$ defined in (\ref{P_cc}).
The 2 lines entering the left side
of the diagram represent the $c$ and $\bar c$ in the ket of the
projection operator. 
The radiative corrections to the matrix element include diagrams in 
which virtual gluons are exchanged 
between the $c$ and $\bar c$ on the right or
between the $c$ and $\bar c$ on the left. They also include
diagrams in which  
real gluons enter the diagram on the left side 
and emerge on the right side with the same momenta and in the 
same color and spin states. 
These contributions
come from higher Fock states in the projection operator 
${\cal P}_{c \bar c',c \bar c}$ defined in (\ref{P_cc}).  
The sum over soft states $S$ in the definition of 
${\cal P}_{c \bar c',c \bar c}$
includes integrals over the momenta of the real gluons and sums
over their spin and color quantum numbers. 
Since $\langle \chi^\dagger \sigma^j T^a \psi \, 
	{\cal P}_{c \bar c',c \bar c} \,
        \psi^\dagger \sigma^i T^a \chi \rangle$
is a vacuum matrix element,  
the only diagrams that are allowed are those for which 
no propagators are cut by a vertical line through the dot.

The coefficient of the factor (\ref{matel-ij})
in (\ref{match-ccbar:2}) is the short-distance coefficient
for the matrix element. The same short-distance coefficient will hold for 
operators defined using the projection 
${\cal P}_{H(\lambda)}$, where $\lambda$ specifies the polarization 
of the quarkonium state $H$.
The color-octet
term in the gluon fragmentation functions at leading order in 
$\alpha_s$ is therefore
\begin{equation}
D_{g \to H(\lambda)}(z) \;=\;
\delta(1-z) {\pi \alpha_s \mu^{2 \epsilon} \over 32 (N-1) m_c^4} \; 
\left( \delta^{ij} - {\hat z}^i {\hat z}^j \right)
\langle \chi^\dagger \sigma^j T^a \psi \, {\cal P}_{H(\lambda)} \,
        \psi^\dagger \sigma^i T^a \chi \rangle .
\label{d-octS:gen}
\end{equation}
The indices $i$ and $j$ range from 1 to $N$.
We will find that, in order to obtain ultraviolet finite results 
at order $\alpha_s^2$, the matrix element in (\ref{d-octS:gen})
will require renormalization.  After that renormalization has been 
carried out, we can take the limit $N \to 3$.  Our final result for 
the order-$\alpha_s$ term in the gluon fragmentation function 
is then
\begin{equation}
D_{g \to H(\lambda)}(z) \;=\;
\delta(1-z) {\pi \alpha_s(m_c) \over 64 m_c^4} \; 
\left( \delta^{ij} - {\hat z}^i {\hat z}^j \right)
\langle \chi^\dagger \sigma^j T^a \psi \, {\cal P}_{H(\lambda)} \,
        \psi^\dagger \sigma^i T^a \chi \rangle^{(\mu)} .
\label{d-octS:3}
\end{equation}
The indices $i$ and $j$ now range  from 1 to 3.
We have set the scale of the running coupling constant
in the short-distance coefficient equal to $m_c$, since
that coefficient is only sensitive to momenta on the 
order of $m_c$ or larger. The superscript $(\mu)$ on the 
matrix element is a reminder that it is a renormalized matrix element
that may depend on a renormalization scale $\mu$.
For a specific quarkonium state, it may be possible to 
simplify the  matrix element in (\ref{d-octS:3}) 
by using rotational symmetry, heavy-quark spin symmetry, 
and the vacuum saturation approximation. For example, 
if we sum over polarizations, we can use 
rotational symmetry to set
\begin{equation}
\sum_\lambda \langle \chi^\dagger \sigma^j T^a \psi \, 
	{\cal P}_{H(\lambda)} \,
        \psi^\dagger \sigma^i T^a \chi \rangle^{(\mu)} 
\;=\;
{1 \over 3} \; \delta^{ij} \; 
\sum_\lambda \langle \chi^\dagger \sigma^k T^a \psi \, 
	{\cal P}_{H(\lambda)} \,
        \psi^\dagger \sigma^k T^a \chi \rangle^{(\mu)} .
\label{rotsym}
\end{equation}

For unpolarized quarkonium states, the calculation of the fragmentation
function can be simplified by using 
rotational symmetry at an earlier stage of the calculation.
Under the rotation group in $N$ dimensions,
${\bf q}$ and ${\bf q}'$ transform as vectors and 
$\xi$, $\eta$, $\xi'$, and $\eta'$ transform as spinors.
Denoting the action of an element $R$ of the rotation group
by $R.{\bf q}$, $R.\xi$, etc.,
the average over the rotation group of a function 
of these vectors and spinors can be defined by 
\begin{equation}
\overline{ f({\bf q},\xi,\eta,{\bf q}',\xi',\eta') } \;\equiv\;
\int dR \; f(R.{\bf q},R.\xi,R.\eta,R.{\bf q}',R.\xi',R.\eta') ,
\label{rotave}
\end{equation}
where $dR$ is the invariant integration element on the group,
normalized so that $\int dR =1$. 
We can average the factor
$( {\cal A}^{\nu a}_{g^* \to c \bar c'(P) + X} )^* 
	{\cal A}^{\mu a}_{g^* \to c \bar c(P) + X}$
on the right side of (\ref{match-frag}) over rotations if we also average the 
projection operator ${\cal P}_{c \bar c',c \bar c}$ on the left side.
Alternatively, we can leave ${\cal P}_{c \bar c',c \bar c}$
unchanged and simply restrict the sum on the left side to
matrix elements for which 
$\langle \chi^\dagger {{\cal K}'}_m^\dagger \psi 
	\psi^\dagger {\cal K}_n \chi \rangle$ is a scalar.

Applying this rotational average to 
the QCD side of the  matching condition in (\ref{match-ccbar:1}), 
it becomes
\begin{equation}
\delta(1-z) {\pi \alpha_s \mu^{2 \epsilon} \over 8 (N-1) m_c^2}  
\left( - g_{\mu\nu} 
	+ {n_\mu {\bar n}_\nu + {\bar n}_\mu n_\nu \over 2} \right) \;
 L^\nu_j \; L^\mu_i \;
\overline{ \eta'^\dagger \sigma^j T^a \xi' \xi^\dagger \sigma^i T^a \eta} \,.
\label{match-ccbar:rot}
\end{equation}
The average over rotations of the spinor factor is 
\begin{equation}
\overline{ \eta'^\dagger \sigma^j T^a \xi' \xi^\dagger \sigma^i T^a \eta }
\;=\; {1 \over N} \delta^{ij} \;
\eta'^\dagger \sigma^k T^a \xi' \xi^\dagger \sigma^k T^a \eta .
\end{equation}
Using the identity (\ref{id:LL}), the 
QCD side of the matching condition in (\ref{match-ccbar:rot}) 
immediately collapses to
\begin{equation}
{\pi \alpha_s \mu^{2 \epsilon} \over 8 N m_c^2} \; \delta(1-z) \;
\eta'^\dagger \sigma^k T^a \xi' \xi^\dagger \sigma^k T^a \eta .
\label{match-ccbar:3}
\end{equation}
The spinor factor is the lowest order expression 
for the matrix element
\begin{equation}
\langle \chi^\dagger \sigma^k T^a \psi \, 
	{\cal P}_{c \bar c',c \bar c} \,
        \psi^\dagger \sigma^k T^a \chi \rangle \Big|_{pNRQCD}
\;\approx\; 4 m_c^2 \;
\eta'^\dagger \sigma^k T^a \xi' \xi^\dagger \sigma^k T^a \eta .
\label{matoct-tree}
\end{equation}
The coefficient of 
this term in (\ref{match-ccbar:3}) is the short-distance 
coefficient for the matrix element.
 Thus the QCD side of the matching condition can be written 
\begin{equation}
{\pi \alpha_s \mu^{2 \epsilon} \over 32 N m_c^4} \; \delta(1-z) \;
\langle \chi^\dagger \sigma^k T^a \psi \, 
	{\cal P}_{c \bar c',c \bar c} \,
        \psi^\dagger \sigma^k T^a \chi \rangle \Big|_{pNRQCD} \,.
\label{match-ccbar:4}
\end{equation}
The fragmentation
function for an unpolarized quarkonium state $H$ is therefore  
\begin{equation} 
D_{g \to H}(z) \;=\;
d^{(\underline{8},^3S_1)}(z) \; 
\langle \chi^\dagger \sigma^k T^a \psi \, {\cal P}_H \,
        \psi^\dagger \sigma^k T^a \chi \rangle^{(\mu)} \,,
\label{Dg-octet}
\end{equation}
 where the short-distance coefficient is 
\begin{equation}
d^{(\underline{8},^3S_1)}(z) \;=\;
{\pi \alpha_s(m_c) \over 96 m_c^4} \delta(1-z) \, .
\label{d-octS}
\end{equation}
In the short-distance coefficient, we have set $N=3$
and set the scale of the running coupling constant equal to $m_c$.
The superscript $(\mu)$ on the matrix element in (\ref{Dg-octet})
indicates that it is a  renormalized matrix element. 
The projection operator ${\cal P}_H$ in the 
matrix element in (\ref{Dg-octet})
is the sum over helicities of the projection operators defined in 
(\ref{P_H}):  ${\cal P}_H = \sum_\lambda {\cal P}_{H(\lambda)}$.
The result (\ref{Dg-octet}) can also be obtained by using the 
identity (\ref{rotsym}) in the unpolarized fragmentation function
(\ref{d-octS:3}).

\section{Leading Color-singlet Terms at order $\alpha_s^2$}

The leading color-singlet terms in the 
gluon fragmentation function $D_{g \to H}(z)$
have short-distance coefficients of order $\alpha_s^2$.  
These terms come from the decay of the virtual gluon through
the process $g^* \to c \bar c g$.  The phase space integrals and the 
energy-momentum-conserving delta function in the matching
condition (\ref{match-frag}) reduce to
\begin{eqnarray}
\int {d^N P \over (2 \pi)^N 2 E_P}
\int {d^N k \over (2 \pi)^N 2 |{\bf k}|} \; 
(2 \pi)^{N+1} \delta^{N+1}(\ell - P - k)
\hspace{1in}
\nonumber \\
\;=\; {1 \over 2} \int_0^1 {d z \over z(1-z)}  
\int {d^{N-1} P_\perp \over (2 \pi)^{N-1}} \;
\delta \left( \ell^2 - {P_\perp^2 + P^2 \over z} 
		- {P_\perp^2 \over 1-z} \right).
\label{phase-space}
\end{eqnarray}
If we only require fragmentation functions that are summed over
the polarizations of the
quarkonium states, the calculation can be simplified by averaging 
both sides of the matching condition over rotations of the 
vectors and spinors that specify the state of the 
$c \bar c$ pair in its rest frame.  The integrand on the 
QCD side of the matching condition (\ref{match-frag})
then reduces to a function
of $P_\perp^2$, and the integration over ${\bf P}_\perp$
can be carried out using the remaining delta function 
in (\ref{phase-space}).  
The QCD side of the matching condition in (\ref{match-frag}) 
reduces to
\begin{eqnarray}
& & { (\sqrt{4 \pi})^{-(N+1)}
 \over 16 \; \Gamma \left({N+1 \over 2}\right) } \;
\int_{4 m_c^2/z}^\infty  {ds \over s^2} \; P_\perp^{N-3}
	\left( - g_{\mu\nu} 
	+ {n_\mu {\bar n}_\nu + {\bar n}_\mu n_\nu \over 2} \right)
\overline{ \left( {\cal A}^{\nu a}_{g^* \to c \bar c'(P) + g} \right)^* 
	{\cal A}^{\mu a}_{g^* \to c \bar c(P) + g} } \Big|_{pQCD} \,,
\label{match-ccbarg}
\end{eqnarray}
where $s = \ell^2$ and $P_\perp^2 = (1-z)(z s - 4 E_q^2)$.
At leading order in $\alpha_s$, the amplitude for the process 
$g^*(l) \to c(p)\; \bar c(\bar{p}) \; g(k) $,  with the $c \bar c$ pair
in a color-singlet state, is
 the sum of the Feynman diagrams in Fig. 3: 
\begin{equation}
{\cal A}^{\mu a}_{g^* \to c \bar c(P) + g}
\;= \; {g_s^2 \mu^{2 \epsilon} \over 6} \; \epsilon^{\nu a}(k) \; \bar u(p)
\left[ \; {\gamma^\mu ( \not \! \bar p \; + \not \! k - m_c) \gamma^\nu
	\over 2 \bar{p} \cdot k}
\;-\; {\gamma^\nu ( \not \! p \; + \not \! k + m_c) \gamma^\mu 
	\over 2 p \cdot k} \; \right] v(\bar p) \;.
\label{M-gccbar} 
\end{equation}
Using the identities for carrying out the nonrelativistic expansion 
of spinor factors given in Appendix A, we expand (\ref{M-gccbar}) 
to linear order in ${\bf q}$: 
\begin{eqnarray}
{\cal A}^{\mu a}_{g^* \to c \bar c(P) + g} 
&\;=\;& {g_s^2 \mu^{2 \epsilon} \over 6 P\cdot k } \; \epsilon^{\nu a}(k)
\Bigg\{   \; m_c \; L^\mu_{\ i} L^\nu_{\ j} (k \cdot L)_k  \;
	\xi^\dagger \{ [ \sigma^i, \sigma^j ], \sigma^k \} \eta
\nonumber \\
&&  \hspace{1in}
\;+\; {2 \over m_c} \Bigg[ 
{4 m_c^2 \over P \cdot k  } \, (k \cdot L)_i
	\left( k^\mu L^\nu_{\ j} + L^\mu_{\ j} \ell^\nu 
		- (k \cdot L)_j \, g^{\mu \nu} \right)
\nonumber \\ 
&& \hspace{1.5in}
\;-\; (k \cdot L)_i ( P^\mu L^\nu_{\ j} - L^\mu_{\ j} P^\nu )
\;+\; (k \cdot L)_j ( P^\mu L^\nu_{\ i} - L^\mu_{\ i} P^\nu )
\nonumber \\ 
&& \hspace{1.5in}
\;+\; P \cdot k \; L^\mu_{\ i} L^\nu_{\ j} 
\;-\; P \cdot \ell \; L^\mu_{\ j} L^\nu_{\ i} 
\; \Bigg] \; \xi^\dagger q^i \sigma^j \eta \; \Bigg\}  .
\label{A-ccbarg}
\end{eqnarray}
In $N>3$ dimensions, the spin matrix 
$\{[\sigma^i,\sigma^j],\sigma^k\}$, 
which is totally antisymmetric in its three indices, is linearly
independent of 1 and $\sigma^i, i=1,\ldots,N$.  In 3 dimensions,
it reduces to the unit matrix multiplied by $-4i \epsilon^{ijk}$.

After inserting the amplitude (\ref{A-ccbarg}) into the 
 QCD side of the 
matching condition (\ref{match-ccbarg}), the spinor factors can be 
simplified by averaging over the rotation group.  The factor 
$\eta'^\dagger q'^k \sigma^l \xi' \xi^\dagger q^i \sigma^j \eta$
can be reduced 
to a linear combination of three independent rotationally-invariant spinor 
factors: 
\begin{eqnarray}
\overline{ \xi'^\dagger \, q'^k \sigma^l \, \eta' 
\eta^\dagger \, q^i \sigma^j \, \xi } 
& = & {1 \over N(N-1)(N+2) } \; 
\nonumber \\
&& 
\times \, \Bigg[\; 
\left( (N+1) 
\delta^{ij} \delta^{kl} - \delta^{ik} \delta^{jl} - \delta^{il} \delta^{jk}  
\right ) 
\eta'^\dagger {\bf q}' \cdot \mbox{\boldmath $\sigma$} \xi' 
	\xi^\dagger {\bf q} \cdot \mbox{\boldmath $\sigma$} \eta 
\nonumber \\
&& \; + \; 
\left( (N+1) 
\delta^{ik} \delta^{jl} - \delta^{ij} \delta^{kl} - \delta^{il} \delta^{jk}  
\right ) 
\, \eta'^\dagger q'^m \sigma^n \xi' \xi^\dagger q^m \sigma^n \eta 
\nonumber \\
&&  \; + \; 
\left( (N+1) 
\delta^{il} \delta^{jk} - \delta^{ij} \delta^{kl} - \delta^{ik} \delta^{jl}  
\right ) 
\, \eta'^\dagger q'^m \sigma^n \xi' \xi^\dagger q^n \sigma^m \eta 
\;\Bigg]. 
\label{spinor-average}
\end{eqnarray}
The spinor factors 
$\eta'^\dagger q'^l \sigma^m \xi' 
	\xi^\dagger \{ [ \sigma^i, \sigma^j ], \sigma^k \} \eta$
and 
$\eta'^\dagger \{ [ \sigma^l, \sigma^m ], \sigma^n \} \xi' 
	\xi^\dagger q^i \sigma^j \eta$
average to zero, while 
$\eta'^\dagger \{ [ \sigma^i, \sigma^j ], \sigma^k \} \xi' 
	\xi^\dagger \{ [ \sigma^i, \sigma^j ], \sigma^k \} \eta$
reduces to a single 
rotationally-invariant spinor factor: 
\begin{equation}
\begin{array}{lcl}
& &
\overline{ 
\eta'^\dagger \{ [ \sigma^l, \sigma^m ], \sigma^n \} \xi' 
	\xi^\dagger \{ [ \sigma^i, \sigma^j ], \sigma^k \} \eta }
\\ [5mm]
& \;= \; &  \displaystyle
 {1 \over N(N-1)(N-2) } 
\;\; \left|\; 
\begin{array}{ccc}
  \delta^{il} ~ & ~ \delta^{jl} ~ & ~ \delta^{kl} \\ 
  \delta^{im} ~ & ~ \delta^{jm} ~ & ~ \delta^{km} \\ 
  \delta^{in} ~ & ~ \delta^{jn} ~ & ~ \delta^{kn} 
\end{array}
\; \right|\; \;
\eta'^\dagger \{ [ \sigma^r, \sigma^s ], \sigma^t \} \xi' 
  \xi^\dagger \{ [ \sigma^r, \sigma^s ], \sigma^t \} \eta .
\end{array}
\label{tensor-average}
\end{equation}
We will find that there are no poles in $N-3$ multipying this spinor factor.
It can therefore be simplified by using the commutation relations 
(\ref{commu}) for $N=3$:
\begin{equation}
\eta'^\dagger \{ [ \sigma^r, \sigma^s ], \sigma^t \} \xi' 
  \xi^\dagger \{ [ \sigma^r, \sigma^s ], \sigma^t \} \eta
\; = \;
-96 \; \eta'^\dagger \xi' \xi^\dagger \eta ,
\qquad N=3 .
\end{equation}

After averaging the spinor factors over the rotation group
using (\ref{spinor-average}) and (\ref{tensor-average}),
all the Cartesian indices of the boost tensors $L^\mu_i$ are contracted
and they can be simplified using the identity (\ref{id:LL})
of Appendix A.  
The factor $\epsilon^{\nu' a}(k)^* \epsilon^{\nu a}(k)$,
summed over colors and polarizations of the real gluon, can be replaced by
$8(- g^{\nu' \nu})$. 
After simplifying all the Lorentz algebra, the QCD side of the 
matching condition (\ref{match-ccbarg}) reduces to integrals
of scalar quantities:
\begin{eqnarray}
 & &
{ (\sqrt{\pi})^{3-N} \alpha_s^2 \mu^{4 \epsilon}
 \over 72 \; \Gamma \left( {N+1 \over 2} \right) 
  N ( N-1 ) ( N+2 )\, m_c^{4+ 2 \epsilon} } 
\int_{(1-z)/z}^\infty  { dx \over x^4 \,( 1 + x)^2 } \; t^{(N-3)/2}
\nonumber \\
& & \hspace{1.5in}
\times \;\Bigg(\; 
m_c^2 \; W_0 (x,z) 
\left( \mbox{$-{1 \over 96}$}
\eta'^\dagger \{ [ \sigma^r, \sigma^s ], \sigma^t \} \xi' 
  \xi^\dagger \{ [ \sigma^r, \sigma^s ], \sigma^t \} \eta \right)
\nonumber \\
& & \hspace{2in}
\; + \; W_1 (x,z)  \; 
\eta'^\dagger {\bf q}' \cdot \mbox{\boldmath $\sigma$} \xi' 
	\xi^\dagger {\bf q} \cdot \mbox{\boldmath $\sigma$} \eta
\; + \; W_2 (x,z)  \; 
	\eta'^\dagger q'^m \sigma^n \xi' \xi^\dagger q^m \sigma^n \eta 
\nonumber \\
& & \hspace{2in}
\; + \; W_3 (x,z)  \; 
	\eta'^\dagger q'^m \sigma^n \xi' \xi^\dagger q^n \sigma^m \eta 
\;\Bigg)\; ,
\label{match-int}
\end{eqnarray}
where $x = (s - 4 m_c^2)/(4 m_c^2)$, $t=(1-z)(zx+z-1)$, and 
\begin{mathletters}
\begin{eqnarray}
W_0 (x,z)  & \; = \; &
6\, (N+2)\, x^2 \left[ - 4 t( 1 + x ) \; + \; (N - 1) x^2 \right] \; ,
\\
W_1 (x,z)  & \; = \; &
- 4 t \left[ 4 N + 4 (N-2) x + (N-9)x^2  + (N-1) x^3 \right]  
\nonumber \\ 
&& \;+\; (N-1) x^2 \left[ 4  - 4 (N + 1) x + (N-1) x^2 \right] \; ,
\\
W_2 (x,z)  & \; = \; &
- 4 t \left[ - 8 - 16 x + ( 2 N^2 +N -17) x^2 + (2 N^2 + N - 9 ) x^3 \right]
\nonumber \\ 
&& \;+\; (N - 1) x^2 \left[  4 (N^2 - 3) +  4 (N^2 - 3) x 
+ (2 N^2 + N - 9) x^2 \right] \; , 
\\
W_3 (x,z)  & \; = \; &
- 4 t \left[ 4 N - 4 ( N^2 - 2 N - 4) x  - (6 N^2 - 5 N - 23) x^2 
	- (2 N^2 - N - 7) x^3 \right]
\nonumber \\ 
&& \;+\; ( N - 1) x^2 \left[ 
    4 - 4 (N^2 - 5) x - (2 N^2 - N - 7) x^2 
	\right] \; .
\end{eqnarray}
\end{mathletters}

Upon integrating (\ref{match-int}) over $x$, we obtain distributions in $z$.
Some of these distributions are infrared divergent in the sense that
the integrals over $z$ diverge as $N \to 3$.  The infrared divergences 
arise from the following integrals:
\begin{eqnarray}
\int_{(1-z)/z}^\infty dx\, {t^{- \epsilon} \over x^2} &\; = \;& 
 \Gamma(1 + \epsilon) \Gamma(1 - \epsilon) \; z \; (1-z)^{-1-2\epsilon}\, ,
\label{IR-int:x2} \\
\int_{(1-z)/z}^\infty dx\, {t^{1- \epsilon} \over x^4} &\; = \;& 
{\Gamma(2 + \epsilon) \Gamma(2 - \epsilon) \over 6}
	\; z^3 \; (1-z)^{-1-2\epsilon} \,.
\label{IR-int:x4}
\end{eqnarray}
The divergences as $\epsilon \to 0$ can be made explicit 
by using the expansion
\begin{eqnarray}
( 1 - z ) ^{ -1 - 2 \epsilon } & = &
-{1 \over 2 \epsilon} \delta ( 1 - z ) + { 1 \over ( 1 - z ) _+ } + \ldots .
\label{expand}
\end{eqnarray}
There is no infrared divergence in the $W_0$ term.
The divergences in the $W_1$ and $W_3$ terms are multiplied
by factors of $\epsilon$, and they give $\delta(1-z)$ terms.  
There is an infrared divergence that survives in the $W_2$ term. 

Integrating over $x$ in (\ref{match-int}) and taking the limit 
$ \epsilon \to 0 $,
the QCD side of the matching condition  reduces to
\begin{eqnarray}
&& 
d^{(\underline{1},^1S_0)}(z) \; 4 m_c^2 \eta'^\dagger \xi' \xi^\dagger \eta 
\nonumber \\
& & 
\; + \; d_1(z) \; 4 m_c^2
\eta'^\dagger {\bf q}' \cdot \mbox{\boldmath $\sigma$} \xi' 
	\xi^\dagger {\bf q} \cdot \mbox{\boldmath $\sigma$} \eta 
\; + \; d_3(z)  \; 4 m_c^2
	\eta'^\dagger q'^m \sigma^n \xi' \xi^\dagger q^n \sigma^m \eta 
\nonumber \\
& & 
\; + \;
\Bigg[ - {\alpha_s^2 \over 324 m_c^6}
	\left( {1 \over \epsilon_{IR}} + \ln(4 \pi \mu^2 ) - \gamma 
	+ {2 \over 3} \right) \delta(1-z)
\nonumber \\
& & \hspace{2in}
	\;+\; d_2(z, \mu ) \Bigg] 4 m_c^2 
	\eta'^\dagger q'^m \sigma^n \xi' \xi^\dagger q^m \sigma^n \eta 
\,.
\label{match-ccbarg:1}
\end{eqnarray}
 The coefficient of the first spinor factor is 
\begin{equation}
d^{(\underline{1},^1S_0)}(z)  \; = \;
{ \alpha_s^2 \over 144 m_c^4 } 
	[  2 (1-z) \ln ( 1- z ) + z (3  - 2 z )  ] \,.
\label{d-singS}
\end{equation}
The remaining three functions in (\ref{match-ccbarg:1}) are 
\begin{mathletters}
\label{d-singP}
\begin{eqnarray}
d_1(z) & = &
{\alpha_s^2 \over 6480 m_c^6 } \; 
[ \delta(1 - z) +  6 (17 - 11 z ) \, \ln (1-z)  +  15 z ( 7  - 2 z) ] \,, 
\label{d-singP:1}
\\
d_2( z, \mu )  & = &
{\alpha_s^2 \over 6480 m_c^6  } \; 
\Bigg[  40 \, { z \over ( 1 - z )_+ }
- 40\, \left( \ln{\mu \over 2 m_c} - {3 \over 20} \right) \delta(1 - z)  
\nonumber \\
 & & \hspace{1in}
\;+\; 36 (2 - z )\,  \ln (1 - z) + 10 z (5 - 4 z )  \Bigg] \,,
\label{d-singP:2}
\\
d_3(z)  & = & 
{\alpha_s^2 \over 6480 m_c^6  } \; 
[\delta (1 - z) +  36 ( 2 - z ) \, \ln (1 - z) + 60 z  ] \,.
\label{d-singP:3}
\end{eqnarray}
\end{mathletters}
The subscript 
 $IR$ on the $\epsilon$ 
in (\ref{match-ccbarg:1})
is a reminder that the pole is of infrared origin.
 The reason for associated the term  
$\ln(4 \pi \mu^2) - \gamma + {2 \over 3}$ with the pole in $\epsilon$
in (\ref{match-ccbarg:1}) will become clear later. 

We now consider the NRQCD side of the 
matching condition (\ref{match-frag}).
The spinor factors in (\ref{match-ccbarg:1}) can be identified as the 
tree-level expressions for 
NRQCD matrix elements,
expanded to linear order in ${\bf q}$ and ${\bf q}'$.
The first matrix element is
\begin{equation}
\langle \chi^\dagger \psi \; 
	{\cal P}_{c \bar c', c \bar c} \;
	\psi^\dagger \chi \rangle \Big|_{pNRQCD}
\; \approx \; 4 m_c^2 \;
\eta'^\dagger \xi' \xi^\dagger \eta \,,
\label{matel-singS}
\end{equation}
and the remaining three are 
\begin{mathletters}
\label{matel-singP}
\begin{eqnarray}
\langle \chi^\dagger (-\mbox{$\frac{i}{2}$} \tensor{\bf D} 
		\cdot \mbox{\boldmath $\sigma$}) \psi \; 
{\cal P}_{c \bar c', c \bar c} \;
\psi^\dagger (-\mbox{$\frac{i}{2}$} \tensor{\bf D} 
		\cdot \mbox{\boldmath $\sigma$}) \chi \rangle 
	\Big|_{pNRQCD}
& \approx & 4 m_c^2 \;
\eta'^\dagger {\bf q}' \cdot \mbox{\boldmath $\sigma$} \xi' 
	\xi^\dagger {\bf q} \cdot \mbox{\boldmath $\sigma$} \eta ,
\label{matel-singP:1}
\\
\langle \chi^\dagger ( -\mbox{$\frac{i}{2}$} \tensor{D})^m \sigma^n \psi \; 
{\cal P}_{c \bar c', c \bar c} \;
\psi^\dagger ( -\mbox{$\frac{i}{2}$} \tensor{D})^m \sigma^n \chi \rangle
	\Big|_{pNRQCD}
&\approx & 4 m_c^2 \;
\eta'^\dagger q'^m \sigma^n \xi' \xi^\dagger q^m \sigma^n \eta ,
\label{matel-singP:2}
\\
\langle \chi^\dagger ( -\mbox{$\frac{i}{2}$} \tensor{D})^m \sigma^n \psi \; 
{\cal P}_{c \bar c', c \bar c} \;
\psi^\dagger ( -\mbox{$\frac{i}{2}$} \tensor{D})^n \sigma^m \chi \rangle
	\Big|_{pNRQCD}
& \approx & 4 m_c^2 \;
\eta'^\dagger q'^m \sigma^n \xi' \xi^\dagger q^n \sigma^m \eta .
\label{matel-singP:3}
\end{eqnarray}
\end{mathletters}
For the matrix elements (\ref{matel-singS}), (\ref{matel-singP:1}),
and (\ref{matel-singP:3}), we can immediately read off 
the short-distance coefficients from (\ref{match-ccbarg:1}).
They are the functions of $z$ given in (\ref{d-singS}),
(\ref{d-singP:1}), and  (\ref{d-singP:3}).
However the coefficient of 
$\eta'^\dagger q'^m \sigma^n \xi' \xi^\dagger q^m \sigma^n \eta$
in (\ref{match-ccbarg:1}) contains an infrared pole in $\epsilon$,
indicating that it is sensitive to long-distance effects.
Since an infrared divergence cannot appear in a short-distance coefficient,
that divergence must be matched by an infrared divergence from some
matrix element on the NRQCD side of the matching condition.
Infrared divergences in NRQCD matrix elements can arise only from 
radiative corrections. 
Since the divergence
in (\ref{match-ccbarg:1}) has a coefficient of order $\alpha_s^2$,
the infrared-divergent NRQCD matrix element  must have a 
short-distance coefficient of order $\alpha_s$.  The only such 
scalar matrix element is the one  
whose short-distance coefficient has already been 
determined in (\ref{match-ccbar:4}).  
Thus the infrared divergence on the NRQCD
side of the matching condition must come from that term.

If the $c \bar c$ pairs are in color-singlet states, the tree level
expression (\ref{matoct-tree}) for the matrix element vanishes
and the leading contribution comes instead from radiative corrections.
Specifically, it comes from the $c \bar c g$ term in the projection
operator  ${\cal P}_{c \bar c',c \bar c}$ defined in (\ref{P_cc}):
\begin{equation}
\sum_{\lambda b} \int {d^Nk \over (2 \pi)^N 2k}
\Big | c({\bf q}',\xi') \bar c(-{\bf q}',\eta') + g({\bf k},\lambda,b) 
\Big \rangle 
\Big \langle 
c({\bf q},\xi) \bar c(-{\bf q},\eta) +  g({\bf k},\lambda,b) \Big | \,.	
\end{equation}
The sum is over the $N-1$ physical polarizations  
and the 8 color states of the real gluon.
The leading contributions to the matrix element are represented by the 
diagrams in Figure 4.  The expression for diagram 4a is
\begin{eqnarray}
& & 
4 g^2 \mu^{2 \epsilon} \;
\eta'^\dagger \sigma^n T^a T^b \xi' 
	\xi^\dagger \sigma^n T^b T^a \eta 
\int {d^Nk \over (2 \pi)^N 2k}
\left( {\bf q} \cdot {\bf q}' 
	- {\bf q} \cdot \hat{\bf k} \; \hat{\bf k} \cdot {\bf q}' \right)
\nonumber \\
& & \hspace{1in} \times
{ 1 \over E_q + k - ({\bf q} + {\bf k})^2/(2 m_c) + i \epsilon} \;
{ 1 \over E_q + k - ({\bf q}' + {\bf k})^2/(2 m_c) + i \epsilon} \,,
\end{eqnarray}
where $E_q = {\bf q}^2/(2 m_c) = ({\bf q}')^2/(2 m_c)$.  
As discussed in Appendix B of  Ref. \cite{B-B-L}, the proper way to 
evaluate the diagram is to first expand out the denominators  
in powers of ${\bf q}/m_c$, ${\bf q}'/m_c$, and ${\bf k}/m_c$, 
and then integrate over ${\bf k}$.  Keeping only terms up to 
linear order in ${\bf q}/m_c$ and ${\bf q}'/m_c$, the diagram reduces to 
\begin{equation}
 8 \pi \alpha_s \mu^{2 \epsilon} \;
\eta'^\dagger \sigma^n T^a T^b \xi'
	\xi^\dagger \sigma^n T^a T^b \eta 
\int {d^Nk \over (2 \pi)^N}
{ {\bf q} \cdot {\bf q}' 
	- {\bf q} \cdot \hat{\bf k} \; \hat{\bf k} \cdot {\bf q}' 
	\over k^3 } \,.
\end{equation}
The integral is both ultraviolet and infrared divergent.  It vanishes in 
dimensional regularization due to a cancellation between an
ultraviolet pole in $\epsilon$ and an infrared pole.  
Making these poles explicit, the diagram can be written
\begin{equation}
{4 \alpha_s \over 3 \pi}
\left( {1 \over \epsilon_{UV}} - {1 \over \epsilon_{IR}} \right)
\eta'^\dagger {q'}^m \sigma^n T^a T^b \xi' 
	\xi^\dagger q^m \sigma^n T^b T^a \eta \,.
\end{equation}
The subscripts $UV$ and $IR$ on $\epsilon$
indicate whether the pole is of ultraviolet or infrared origin.
We have set $N=3$ in the prefactor, since any finite terms obtained by 
expanding the prefactor in powers of $\epsilon$ will cancel. 
The effect of the other 3 diagrams is simply to symmetrize both 
of the products of color matrices $T^a T^b$.  Since the  
projector ${\cal P}_{c \bar c',c \bar c}$ requires the asymptotic
$c \bar c$ pairs to be 
in color-singlet states, we can replace $\{T^a,T^b\}$
by $\delta^{ab}/3$.  The final result for the 
matrix element is 
\begin{equation} 
\langle \chi^\dagger \sigma^k T^a \psi \, {\cal P}_{c \bar c',c \bar c} \,
        \psi^\dagger \sigma^k T^a \chi \rangle \Big|_{pNRQCD}
\;=\; {32 \alpha_s \over 27 \pi} 
\left( {1 \over \epsilon_{UV}} - {1 \over \epsilon_{IR}} \right)
\eta'^\dagger {q'}^m \sigma^n \xi' 
	\xi^\dagger q^m \sigma^n \eta \,.
\label{matel-bare}
\end{equation}
After multiplying by the short-distance coefficient in 
(\ref{match-ccbar:4}), we find that the infrared pole in $\epsilon$
matches the one on the QCD side of the matching condition, 
which is given in (\ref{match-ccbarg:1}).

After taking into account the matrix element (\ref{matel-bare})
on the NRQCD side of the matching condition, the short-distance 
coefficient of the matrix element (\ref{matel-singP:2})
can be read off from (\ref{match-ccbarg:1}).
The net effect of taking into account the matrix element
(\ref{matel-bare}) is simply to change the infrared pole in $\epsilon$
into an ultraviolet pole.  The short-distance coefficient of the 
matrix element (\ref{matel-singP:2})
therefore contains an ultraviolet divergence.
This divergence must be removed by operator renormalization in NRQCD.
As is evident from the ultraviolet pole in $\epsilon$ in (\ref{matel-bare}),
it is the matrix element
$\langle \chi^\dagger \sigma^k T^a \psi {\cal P}_{c \bar c',c \bar c}
        \psi^\dagger \sigma^k T^a \chi \rangle$ 
that requires renormalization.
In the $\overline{MS}$ renormalization scheme,  
the relation betweeen the matrix element of the bare operator 
and matrix elements of renormalized operators is 
\begin{eqnarray}
&&
\langle \chi^\dagger \sigma^k T^a \psi \, {\cal P}_{c \bar c',c \bar c} \,
        \psi^\dagger \sigma^k T^a \chi \rangle
\;=\; \mu^{- 4 \epsilon} \Bigg(
\langle \chi^\dagger \sigma^k T^a \psi \, {\cal P}_{c \bar c',c \bar c} \,
        \psi^\dagger \sigma^k T^a \chi \rangle^{(\mu)}  
\nonumber \\
 & & \hspace{0.5in}
\;+\; {8 \alpha_s \over 27 \pi m_c^2} 
\left( {1 \over \epsilon_{UV}} + \ln(4 \pi) - \gamma \right)
\langle \chi^\dagger ( -\mbox{$\frac{i}{2}$} \tensor{D})^m \sigma^n \psi \, 
{\cal P}_{c \bar c', c \bar c} \,
\psi^\dagger ( -\mbox{$\frac{i}{2}$} \tensor{D})^m 
	\sigma^n \chi \rangle^{(\mu)} \Bigg) \, .
\label{matel-ren:def}
\end{eqnarray}
The superscripts $(\mu)$ on the matrix elements on the right
side indicate that they are renormalized matrix elements with 
renormalization scale $\mu$.  
We will suppress this superscript on P-wave matrix elements,
since they do not require any renormalization at this order in $\alpha_s$.
The  fermion field operators in the bare matrix element on the left side 
of (\ref{matel-ren:def}) have dimension $N/2$.  
The fermion field operators in the renormalized 
matrix elements on the right side have dimension $3/2$.
The factor of $\mu^{-4 \epsilon}$ on the right side of (\ref{matel-ren:def})
compensates for the difference between the dimensions of the two sides. 
Solving (\ref{matel-ren:def}) for 
$\langle \chi^\dagger \sigma^k T^a \psi  {\cal P}_{c \bar c',c \bar c}
        \psi^\dagger \sigma^k T^a \chi \rangle^{(\mu)}$ and using
(\ref{matel-bare}) and (\ref{matel-singP:2}),        
we find that the renormalized matrix element, with dimensional 
regularization as the infrared cutoff, is
\begin{eqnarray} 
&&
\langle \chi^\dagger \sigma^k T^a \psi \, {\cal P}_{c \bar c',c \bar c} \,
        \psi^\dagger \sigma^k T^a \chi \rangle^{(\mu)} \Big|_{pNRQCD}
\nonumber \\
&& \hspace{1.5in}
\;=\; - {32 \alpha_s \over 27 \pi}
\left( {1 \over \epsilon_{IR}}  + \ln(4 \pi) - \gamma \right)
\eta'^\dagger {q'}^m \sigma^n \xi' 
	\xi^\dagger q^m \sigma^n \eta \,.
\label{matel-ren}
\end{eqnarray}
Multiplying (\ref{matel-ren:def}) by the short-distance 
coefficient in (\ref{match-ccbar:4}), we find that the contribution 
from the renormalized matrix element to the 
NRQCD side of the matching condition is
\begin{eqnarray}
&&
{\pi \alpha_s \mu^{2 \epsilon} \over 32 N m_c^4} \; \delta (1-z) \; 
\langle \chi^\dagger \sigma^k T^a \psi \, {\cal P}_{c \bar c',c \bar c} \,
        \psi^\dagger \sigma^k T^a \chi \rangle^{(\mu)} \Big|_{pNRQCD}
\nonumber \\
&& \hspace{.5in}
\;=\; - {\alpha_s^2 \over 324 m_c^6}
\left({1 \over \epsilon_{UV}} + \ln(4 \pi \mu^2) 
	- \gamma + {2 \over 3} \right) \delta(1-z) \;
4 m_c^2 \eta'^\dagger {q'}^m \sigma^n \xi' 
	\xi^\dagger q^m \sigma^n \eta \,.
\label{match-NRQCD}
\end{eqnarray}
This term on the NRQCD side of the matching condition 
matches the part of the 
$\eta'^\dagger {q'}^m \sigma^n \xi' \xi^\dagger q^m \sigma^n \eta$
term on the QCD side
that contains the pole in $\epsilon$.  Subtracting this term from
(\ref{match-ccbarg:1}), we find that the  
short-distance coefficient of the matrix element (\ref{matel-singP:2})
is the function $d_2(z,\mu)$ given in (\ref{d-singP:2}).

Note that the calculation of the radiative correction to the 
NRQCD matrix element in (\ref{matel-bare}) was necessary only to 
determine the coefficient of the counterterm in (\ref{matel-ren:def}).
The radiative correction itself vanishes if we identify 
$\epsilon_{UV} = \epsilon_{IR}$.  Its only effect is to transform the 
ultraviolet pole in the coefficient of the counterterm into an infrared pole 
on the NRQCD side of the matching condition.

Combining the color-octet term in (\ref{Dg-octet}) with
the color-singlet terms determined above, we obtain
a general expression for the gluon fragmentation 
function of an unpolarized quarkonium state $H$:
\begin{eqnarray}
D_{g \to H} (z) 
& = &  d^{(\underline{8},^3S_1)}(z) \; 
\langle \chi^\dagger \sigma^n T^a \psi \; 
	{\cal P}_H \; \psi^\dagger \sigma^n T^a \chi \rangle^{(\mu)}
\;+\; d^{(\underline{1},^1S_0)}(z) \; 
\langle \chi^\dagger \psi \; 
	{\cal P}_H \; \psi^\dagger \chi \rangle 
\nonumber \\
 & & \hspace{1.in}
\; + \; d_1(z) \; 
\langle \chi^\dagger (-\mbox{$\frac{i}{2}$} \tensor{\bf D} 
		\cdot \mbox{\boldmath $\sigma$}) \psi \; 
	{\cal P}_H \;
\psi^\dagger (-\mbox{$\frac{i}{2}$} \tensor{\bf D} 
		\cdot \mbox{\boldmath $\sigma$}) \chi \rangle 
\nonumber \\
 & & \hspace{1.in}
\; + \; d_2(z,\mu) \; 
\langle \chi^\dagger (-\mbox{$\frac{i}{2}$} \tensor{D})^m \sigma^n \psi \; 
	{\cal P}_H \;
\psi^\dagger (-\mbox{$\frac{i}{2}$} \tensor{D})^m \sigma^n \chi \rangle
\nonumber \\
 & & \hspace{1.in}
\; + \; d_3(z) \; 
\langle \chi^\dagger (-\mbox{$\frac{i}{2}$} \tensor{D})^m \sigma^n \psi \; 
	{\cal P}_H \; 
\psi^\dagger (-\mbox{$\frac{i}{2}$} \tensor{D})^n \sigma^m \chi \rangle 
\, ,
\label{Dg-gen}
\end{eqnarray}
where the short-distance coefficients are given in (\ref{d-octS}),
(\ref{d-singS}), and (\ref{d-singP}).
This expression for the fragmentation function
includes all those color-octet matrix elements with short-distance 
coefficient of order $\alpha_s$
and all those color-singlet matrix elements with short-distance 
coefficients of order $\alpha_s^2$ that involve at most one 
derivative acting on
$\chi^\dagger$ and $\psi$ and  at most one derivative acting on
$\psi^\dagger$ and $\chi$. 

The $\mu$-dependence of the 
coefficient $d_2(z,\mu)$ in (\ref{Dg-gen})
is cancelled by the $\mu$-dependence 
of the color-octet matrix element.
The renormalization group equation that determines the 
$\mu$-dependence of that matrix element is obtained by differentiating
 (\ref{matel-ren:def}) with respect to $\mu$.  
We use the fact that the bare matrix element on the left side is 
independent of $\mu$ and that the bare coupling constant, 
which at this order is $\alpha_s \mu^{2 \epsilon}$, 
is also independent of $\mu$.  Replacing
${\cal P}_{c \bar c',c \bar c}$ by the projector ${\cal P}_H$,
we find 
\begin{equation}
\mu {d \ \over d \mu}
\langle \chi^\dagger \sigma^k T^a \psi \, {\cal P}_H \,
        \psi^\dagger \sigma^k T^a \chi \rangle^{(\mu)} 
\;=\; {16 \over 27 \pi m_c^2}  \alpha_s(\mu)
\langle \chi^\dagger ( -\mbox{$\frac{i}{2}$} \tensor{D})^m \sigma^n \psi \; 
	{\cal P}_H \;
\psi^\dagger ( -\mbox{$\frac{i}{2}$} \tensor{D})^m \sigma^n \chi \rangle
\, .
\label{matel-rgeq}
\end{equation}
The solution to this renormalization group equation at leading order 
in $\alpha_s$ is  
\begin{eqnarray}
\langle \chi^\dagger \sigma^k T^a \psi \, {\cal P}_H \,
        \psi^\dagger \sigma^k T^a \chi \rangle^{(\mu)} 
&\;=\;& 
\langle \chi^\dagger \sigma^k T^a \psi \, {\cal P}_H \,
        \psi^\dagger \sigma^k T^a \chi \rangle^{(\mu_0)} 
\nonumber \\
&& \hspace{-1in}
\;-\; {32  \over 9 (33 - 2 n_f) m_c^2} 
\log \left({\alpha_s(\mu) \over \alpha_s(\mu_0)} \right)
\langle \chi^\dagger ( -\mbox{$\frac{i}{2}$} \tensor{D})^m \sigma^n \psi \; 
	{\cal P}_H \;
\psi^\dagger ( -\mbox{$\frac{i}{2}$} \tensor{D})^m \sigma^n \chi \rangle
\, ,
\label{matel-rgsol}
\end{eqnarray}
where $n_f$ is the number of flavors of light quarks ($n_f=3$ for charmonium). 
 
\section{Gluon Fragmentation Functions }

The relative importance of the various terms in 
the fragmentation function (\ref{Dg-gen})
depends on the quarkonium state.  The magnitude of a particular term
is determined by the order in $\alpha_s$ of its short-distance coefficient
and by the scaling of the matrix element with $v$,
which is given by the velocity-scaling rules of NRQCD \cite{B-B-L}. 
Below, we apply this general 
fragmentation formula to specific S-wave states and P-wave states.

\subsection{Spin-singlet S-wave states}

The dominant Fock state of the $\eta_c$ consists of a $c \bar c$ pair in a 
color-singlet ${}^1S_0$ state.  
The dominant matrix element is therefore 
$\langle \chi^\dagger \psi {\cal P}_{\eta_c} \psi^\dagger \chi \rangle$, 
which scales as $v^3$. It has a short-distance coefficient 
of order $\alpha_s^2$, and therefore contributes to the gluon 
fragmentation function at order $\alpha_s^2 v^3$.
Since the dominant Fock state can be reached
from a color-octet ${}^3S_1$ state through a chromomagnetic
dipole transition, the matrix element
$\langle \chi^\dagger \sigma^k T^a \psi {\cal P}_{\eta_c}
        \psi^\dagger \sigma^k T^a \chi \rangle$
is suppressed  by $v^3$ relative to the dominant matrix element.
It contributes to $D_{g\to \psi}(z)$ at order  $\alpha_s v^6$.
The color-singlet P-wave matrix elements in (\ref{Dg-gen}) 
are all suppressed by $v^7$.  Their contributions are therefore of order
$\alpha_s^2 v^{10}$, and are expected to be negligible.

The most important term in the gluon fragmentation function
for the $\eta_c$ is the color-singlet $^1S_0$ term.
Keeping only this term, the fragmentation function reduces to
\begin{equation}
D_{g \to \eta_c}(z)  \;= \; 
{\alpha_s^2(m_c) \over 144 m_c^4 }
\; \left [ 3z - 2 z^2 + 2 (1-z) \ln ( 1- z ) \right ]
\langle \chi^\dagger \psi \, {\cal P}_{\eta_c} \,
        \psi^\dagger \chi \rangle \,.
\label{Dg-eta}
\end{equation}
 Since the short-distance coefficient is dominated by 
contributions from large momenta of order $m_c$ or larger,
we have set the scale of the running coupling constant equal to $m_c$. 
The standard NRQCD matrix elements 
$\langle {\cal O}_n(^{2S+1}L_J) \rangle$ introduced in Ref. \cite{B-B-L} 
were defined using a projection operator analogous to ${\cal P}_H$
in (\ref{P_H}),
except that the  states have the standard nonrelativistic normalization.
The relation between our matrix elements and those defined in 
Ref. \cite{B-B-L} 
is discussed in Appendix B of \cite{Braaten-Chen}.  At leading 
order in $v^2$, they  differ simply by a normalization factor.
For the matrix element in (\ref{Dg-eta}), the relation is 
\begin{equation}
\langle \chi^\dagger \psi \, {\cal P}_{\eta_c} \,
        \psi^\dagger \chi \rangle 
\; \approx \;
4 m_c \; \langle {\cal O}_1^{\eta_c} ({}^1S_0) \rangle \,.
\end{equation}
The term (\ref{Dg-eta}) in the fragmentation function was first 
calculated by Braaten and Yuan \cite{Braaten-Yuan:Swave}.
The next most important term should be the color-octet 
${}^3S_1$ term in (\ref{Dg-gen}).

\subsection{Spin-triplet S-wave states}

The dominant Fock state of the $J/\psi$ consists of a $c \bar c$ pair in a 
color-singlet ${}^3S_1$ state.  
The dominant matrix element is therefore 
$\langle \chi^\dagger \sigma^k \psi {\cal P}_{\psi}
        \psi^\dagger \sigma^k \chi \rangle$, 
which scales as $v^3$. 
Since the dominant Fock state can be reached
from a color-octet ${}^3S_1$ state through a double chromoelectric
dipole transition, the matrix element
$\langle \chi^\dagger \sigma^k T^a \psi {\cal P}_{\psi}
        \psi^\dagger \sigma^k T^a \chi \rangle$
is suppressed  by $v^4$ relative to the dominant matrix element.
The matrix element
$\langle \chi^\dagger \psi  
	{\cal P}_{\psi} \psi^\dagger \chi \rangle$ 
is suppressed by $v^7$
and the color-singlet P-wave matrix elements in (\ref{Dg-gen}) 
are all suppressed by $v^8$.  Their contributions are therefore 
expected to be negligible.

Of the terms that we have calculated, the most important is 
the color-octet $^3S_1$ term.  Keeping only this term, 
the fragmentation function reduces to
\begin{equation}
D_{g \to \psi}(z) \;=\;
\delta(1-z) {\pi \alpha_s(m_c) \over 96   m_c^4} \; 
\langle \chi^\dagger \sigma^i T^a \psi \, {\cal P}_\psi \,
        \psi^\dagger \sigma^i T^a \chi \rangle  .
\label{Dg-psi}
\end{equation}
 Since the short-distance coefficient is dominated by 
contributions from large momenta of order $m_c$ or larger,
we have set the scale of the running coupling constant equal to $m_c$. 
Up to corrections of relative order $v^2$, the matrix element in 
(\ref{Dg-psi}) differs from the standard NRQCD matrix element 
introduced in Ref. \cite{B-B-L} 
only by a normalization factor: 
\begin{equation}
\langle \chi^\dagger \sigma^i T^a \psi \, {\cal P}_\psi \,
        \psi^\dagger \sigma^i T^a \chi \rangle 
\;\approx\;
4 m_c \; \langle {\cal O}_8^{\psi} ({}^3S_1) \rangle  \, .
\end{equation}
The result (\ref{Dg-psi}) was first given by 
Braaten and Fleming \cite{Braaten-Fleming:Swave}.
This contribution to the fragmentation function is of order
$\alpha_s v^7$.
If we assume that the effect of a suppression factor of  $\alpha_s$ 
is comparable to that of a suppression factor of $v^2$, 
then there is one other term in the fragmentation function
that could be equally important.
The color-singlet $^3S_1$ term contributes at order 
$\alpha_s^3 v^3$.  It has been calculated in Ref. \cite{Braaten-Yuan:psiCSM}.

\subsection{Spin-singlet P-wave states}

The dominant Fock state of the $h_c$ consists of a $c \bar c$ pair in a 
color-singlet ${}^1P_1$ state.  
According to the velocity-scaling rules,
the largest matrix elements are 
$\langle \chi^\dagger (-\mbox{$\frac{i}{2}$} \tensor{\bf D})^m \psi
	{\cal P}_{h_c} 
\psi^\dagger (-\mbox{$\frac{i}{2}$} \tensor{\bf D})^m \chi \rangle$
and  
$\langle \chi^\dagger T^a \psi {\cal P}_{h_c} \psi^\dagger T^a \chi \rangle$, 
both of which scale  as $v^5$.
The reason that they are comparable in magnitude is that one 
is suppressed by a factor of $v^2$ from the covariant derivatives, while
the other is suppressed by $v^2$  because of the chromoelectric dipole 
transition required to reach the dominant Fock state from a color-octet 
$^1S_0$ state. The color-octet $^3S_1$ and color-singlet $^1S_0$
matrix elements in (\ref{Dg-gen}) 
contribute at orders $\alpha_s v^8$ and $\alpha_s^2 v^9$, 
respectively. 
The color-singlet P-wave matrix elements in (\ref{Dg-gen}) 
contribute at order $\alpha_s^2 v^{12}$ and should be negligible.
If we assume that the effect of a suppression factor of  $\alpha_s$ 
is comparable to that of a suppression factor of $v^2$, then the most 
important term is the color-octet $^1S_0$ term, which contributes 
at order $\alpha_s^2 v^5$.  This term has not yet been calculated.

\subsection{Spin-triplet P-wave states}

The dominant Fock state of the $\chi_{cJ}$ consists of a $c \bar c$ pair in a 
color-singlet ${}^3P_J$ state.  
According to the velocity-scaling rules,
the largest matrix elements are 
$\langle \chi^\dagger \sigma^i T^a \psi {\cal P}_{\chi_{cJ}} 
	\psi^\dagger \sigma^i T^a \chi \rangle$
	and the color-singlet P-wave matrix elements in (\ref{Dg-gen}), 
all of which scale  as $v^5$.  The color-octet $^3S_1$ matrix element
contributes to the gluon fragmentation function at order $\alpha_s v^5$,
while  the color-singlet P-wave matrix elements contribute at order
$\alpha_s^2 v^5$.  The matrix element
$\langle \chi^\dagger \psi {\cal P}_{\chi_{cJ}} \psi^\dagger \chi \rangle$ 
contributes at order  $\alpha_s^2 v^8$ and should be negligible.
We will therefore consider only the color-octet $^3S_1$ term and the 
color-singlet P-wave terms in the fragmentation function.
 
The symmetries of NRQCD can be used to reduce the three 
color-octet S-wave matrix elements
$\langle \chi^\dagger \sigma^i T^a \psi {\cal P}_{\chi_{cJ}} 
	\psi^\dagger \sigma^i T^a \chi \rangle$
to one independent matrix element, which we take to be the $J=0$ one.
These symmetries are rotational symmetry and heavy-quark spin symmetry.  
Spin symmetry is an approximate symmetry with corrections that are of
relative order $v^2$.  It implies that the state
$\chi_{cJ}(\lambda)$, which is an eigenstate of ${\bf J}^2$ and $J_z$,
can be expressed in the form
\begin{equation}
\Big | \chi_{cJ}(\lambda) \Big \rangle 
\;\approx\; \sum_{l_z s_z} \langle 1 l_z; 1 s_z | J \lambda \rangle
\Big | \chi_c(l_z s_z) \Big \rangle \,,
\label{chi}
\end{equation}
where  $\chi_c(l_z s_z)$ is an eigenstate of $L_z$ and $S_z$.
Using the definition (\ref{P_H}) of the projection operator ${\cal P}_H$,
the color-singlet $^3S_1$ matrix element can be written
\begin{eqnarray}
\langle \chi^\dagger \sigma^i T^a \psi \, {\cal P}_{\chi_{cJ}} \,
	\psi^\dagger \sigma^i T^a \chi \rangle 
&\approx& \sum_\lambda \sum_{l_z s_z} \sum_{l_z' s_z'}
	\langle 1 l_z; 1 s_z | J \lambda \rangle 
	\langle J \lambda | 1 l_z'; 1 s_z' \rangle 
\nonumber \\
&& \times 
\sum_S \Big \langle 0 \Big | \chi^\dagger \sigma^i T^a \psi
	\Big | \chi_c(l_z s_z)+ S \Big \rangle 
	\Big \langle \chi_c(l_z' s_z') + S \Big |
	\psi^\dagger \sigma^i T^a \chi \Big | 0 \Big \rangle \,.
\label{octS-SS}
\end{eqnarray}
Spin symmetry also implies that the two matrix elements in (\ref{octS-SS})
are proportional to $U_{s_z i}$ and $U^\dagger_{i s_z'}$, where
$U_{mi}$ is the unitary $3 \times 3$ matrix that transforms vectors from the 
Cartesian basis to the spherical basis.
Finally, rotational symmetry implies that the product of the two 
matrix elements in (\ref{octS-SS}) summed over soft states $S$
must be proportional to $\delta_{l_z l_z'}$.
Using the orthogonality relations of the Clebsch-Gordan coefficients,
the equation (\ref{octS-SS}) can be reduced to 
\begin{equation}
\langle \chi^\dagger \sigma^i T^a \psi \, {\cal P}_{\chi_{cJ}} \,
	\psi^\dagger \sigma^i T^a \chi \rangle 
\;\approx\;  
(2J+1) \; \langle \chi^\dagger \sigma^i T^a \psi \, {\cal P}_{\chi_{c0}} \,
	\psi^\dagger \sigma^i T^a \chi \rangle\,.
\label{octS-id}
\end{equation}
These relations hold up to corrections of relative order $v^2$.

The color-singlet P-wave matrix elements in (\ref{Dg-gen}) can also 
be reduced to a single independent matrix element, which we choose to be
$\langle \chi^\dagger (-\mbox{$\frac{i}{2}$} \tensor{\bf D} 
		\cdot \mbox{\boldmath $\sigma$}) \psi 
	{\cal P}_{\chi_{c0}}
\psi^\dagger (-\mbox{$\frac{i}{2}$} \tensor{\bf D} 
		\cdot \mbox{\boldmath $\sigma$}) \chi \rangle$.
Using the expression (\ref{chi}) for the 
$\chi_{cJ}$
states and the definition (\ref{P_H}), we can write
\begin{eqnarray}
\langle \chi^\dagger (-\mbox{$\frac{i}{2}$} \tensor{\bf D})^i \sigma^j \psi 
	{\cal P}_{\chi_{cJ}}
\psi^\dagger (-\mbox{$\frac{i}{2}$} \tensor{\bf D})^m \sigma^n \chi \rangle
\;\approx\; \sum_\lambda \sum_{l_z s_z} \sum_{l_z' s_z'}
	\langle 1 l_z; 1 s_z | J \lambda \rangle 
	\langle J \lambda | 1 l_z'; 1 s_z' \rangle 
\nonumber \\
\times \sum_S \Big \langle 0  \Big | 
	\chi^\dagger (-\mbox{$\frac{i}{2}$} \tensor{\bf D})^i \sigma^j \psi
	\Big | \chi_c(l_z s_z)+ S \Big \rangle 
	\Big \langle \chi_c(l_z' s_z') + S \Big |
	\psi^\dagger (-\mbox{$\frac{i}{2}$} \tensor{\bf D})^m \sigma^n \chi
	\Big | 0 \Big \rangle \,.
\label{singP-SS}
\end{eqnarray}
Spin symmetry implies that the two matrix elements in (\ref{singP-SS})
are proportional to $U_{s_z j}$ and $U^\dagger_{n s_z'}$.
Since the operators create and annihilate $c \bar c$ pairs in the
dominant Fock state of $\chi_c$, we can use the vacuum-saturation 
approximation, which is accurate up to corrections of relative order $v^4$.
Keeping only the vacuum term in the sum over $S$, 
the matrix elements reduce to
$\langle 0 |
	\chi^\dagger (-\mbox{$\frac{i}{2}$} \tensor{\bf D})^i \sigma^j \psi
	| \chi_c(l_z s_z) \rangle$ and 
$\langle \chi_c(l_z' s_z') |
	\psi^\dagger (-\mbox{$\frac{i}{2}$} \tensor{\bf D})^m \sigma^n \chi
	| 0 \rangle$.
By rotational symmetry, these must be proportional to 
$U_{l_z i}$ and $U^\dagger_{m l_z'}$.  Thus the tensorial structure
of the matrix element (\ref{singP-SS}) is completely determined.
The proportionality constant can be deduced by taking the special case 
$J=0$, $i=j$ and $m=n$.  The resulting formula is
\begin{eqnarray}
\langle \chi^\dagger (-\mbox{$\frac{i}{2}$} \tensor{\bf D})^i \sigma^j \psi 
	{\cal P}_{\chi_{cJ}}
\psi^\dagger (-\mbox{$\frac{i}{2}$} \tensor{\bf D})^m \sigma^n \chi \rangle
\;\approx\; {1 \over 3} \;
\langle \chi^\dagger  (-\mbox{$\frac{i}{2}$} \tensor{\bf D} 
		\cdot \mbox{\boldmath $\sigma$}) \psi 
	{\cal P}_{\chi_{c0}}
\psi^\dagger  (-\mbox{$\frac{i}{2}$} \tensor{\bf D} 
		\cdot \mbox{\boldmath $\sigma$}) \chi \rangle
\nonumber \\
\;\times\; 
\sum_\lambda \sum_{l_z s_z} \sum_{l_z' s_z'}
	\langle 1 l_z; 1 s_z | J \lambda \rangle 
	\langle J \lambda | 1 l_z'; 1 s_z' \rangle 
	U_{s_z j} U^\dagger_{n s_z'} U_{l_z i} U^\dagger_{m l_z'} \,.
\label{singP-tensor}
\end{eqnarray}
The scalar combinations of these matrix elements can be simplified
by using the orthogonality relations of Clebsch-Gordan coefficients 
together with the identity 
\begin{equation}
\left( U U^t \right)_{m_1 m_2} 
\;=\; - \sqrt{3} \; \langle 1 m_1; 1 m_2 | 0 0 \rangle \,.
\label{UUt}
\end{equation}
The resulting formulas are
\begin{mathletters}
\label{singP-scalar}
\begin{eqnarray}
\langle \chi^\dagger (-\mbox{$\frac{i}{2}$} \tensor{\bf D} 
		\cdot \mbox{\boldmath $\sigma$}) \psi 
	{\cal P}_{\chi_{cJ}}
\psi^\dagger (-\mbox{$\frac{i}{2}$} \tensor{\bf D} 
		\cdot \mbox{\boldmath $\sigma$}) \chi \rangle
& \;\approx\; & \delta_{J0} \;
\langle \chi^\dagger  (-\mbox{$\frac{i}{2}$} \tensor{\bf D} 
		\cdot \mbox{\boldmath $\sigma$}) \psi 
	{\cal P}_{\chi_{c0}}
\psi^\dagger  (-\mbox{$\frac{i}{2}$} \tensor{\bf D} 
		\cdot \mbox{\boldmath $\sigma$}) \chi \rangle \,,
\\
\langle \chi^\dagger (-\mbox{$\frac{i}{2}$} \tensor{\bf D})^m \sigma^n \psi 
	{\cal P}_{\chi_{cJ}}
\psi^\dagger (-\mbox{$\frac{i}{2}$} \tensor{\bf D})^m \sigma^n \chi \rangle
& \;\approx\; & {2J+1 \over 3} \;
\langle \chi^\dagger  (-\mbox{$\frac{i}{2}$} \tensor{\bf D} 
		\cdot \mbox{\boldmath $\sigma$}) \psi 
	{\cal P}_{\chi_{c0}}
\psi^\dagger  (-\mbox{$\frac{i}{2}$} \tensor{\bf D} 
		\cdot \mbox{\boldmath $\sigma$}) \chi \rangle \,,
\\
\langle \chi^\dagger (-\mbox{$\frac{i}{2}$} \tensor{\bf D})^m \sigma^n \psi 
	{\cal P}_{\chi_{cJ}}
\psi^\dagger (-\mbox{$\frac{i}{2}$} \tensor{\bf D})^n \sigma^m \chi \rangle
&&
\nonumber \\
&& \hspace{-1in}
 \;\approx\; (-1)^J {2J+1 \over 3} \;
\langle \chi^\dagger  (-\mbox{$\frac{i}{2}$} \tensor{\bf D} 
		\cdot \mbox{\boldmath $\sigma$}) \psi 
	{\cal P}_{\chi_{c0}}
\psi^\dagger  (-\mbox{$\frac{i}{2}$} \tensor{\bf D} 
		\cdot \mbox{\boldmath $\sigma$}) \chi \rangle \,.
\end{eqnarray}
\end{mathletters}

Using the relations (\ref{octS-id}) and (\ref{singP-scalar}),
the dominant terms in the gluon fragmentation function for the 
$\chi_{cJ}$ reduce to
\begin{eqnarray}
D_{g \to \chi_{cJ}} (z) 
& = &  (2J+1) \; d^{(\underline{8},^3S_1)}(z) \; 
\langle \chi^\dagger \sigma^n T^a \psi \; 
	{\cal P}_{\chi_{c0}} \; \psi^\dagger \sigma^n T^a \chi \rangle^{(\mu)}
\nonumber \\
 & & \hspace{.5in}
\;+\; d^{(\underline{1},^3P_J)}(z,\mu) \; 
\langle  \chi^\dagger (-\mbox{$\frac{i}{2}$} \tensor{\bf D} 
		\cdot \mbox{\boldmath $\sigma$}) \psi
	\; {\cal P}_{\chi_0 } \;
	\psi^\dagger (-\mbox{$\frac{i}{2}$} \tensor{\bf D} 
		\cdot \mbox{\boldmath $\sigma$}) \chi
\rangle  \,,
\label{Dg-chi}
\end{eqnarray}
where $d^{(\underline{8},^3S_1)}(z)$ is given in (\ref{d-octS})
and $d^{(\underline{1},^3P_J)}(z,\mu)$ is a  linear combination
of the functions in (\ref{d-singP}): 
\begin{equation}
d^{(\underline{1},^3P_J)}(z,\mu)
\;=\; \delta_{J0} \; d_1(z)
\;+\; {2J+1 \over 3} \;  d_2(z,\mu)
\;+\; (-1)^J {2J+1 \over 3} \; d_3(z) \,.
\label{d-3P}
\end{equation}
More explicitly, these coefficients are 
\begin{eqnarray}
d^{(\underline{1},^3P_J)}(z,\mu)
&=& {\alpha_s^2(m_c) \over 486 m_c^6}
\Bigg[\, (2J+1) {z \over (1-z)_+}
\nonumber \\
&& \hspace{.5in}
\;+\; \left( Q_J - (2J + 1) \log{\mu \over 2 m_c} \right) \delta(1-z)
	\; + \; P_J(z) \Bigg] \; ,
\label{D1-J} 
\end{eqnarray}
where the numbers $Q_J$ are
\begin{equation}
Q_0 \;=\;  { 1 \over 4 } \;, \quad
Q_1 \;=\;  { 3 \over 8  } \;, \quad
Q_2 \;=\;  { 7 \over 8  } \;, \quad
\label{qJ} 
\end{equation}
and the functions $P_J(z)$ are
\begin{eqnarray}
P_0(z) &=& {z (85 -  26 z) \over 8}
	\; + \; {9 (5 - 3z) \over 4} \; \log(1-z) \; ,
\label{pzero} 
\\
P_1(z) &=& - {3 z (1 + 4 z) \over 4} \; ,
\label{pone} 
\\
P_2(z) &=& {5 z (11 - 4z) \over 4}
	\; + \; { 9 } (2-z) \; \log(1-z) \; .
\label{ptwo} 
\end{eqnarray}
The matrix elements in (\ref{Dg-chi}) are related to the standard
matrix elements defined in Ref. \cite{B-B-L} by
\begin{eqnarray}
\langle \chi^\dagger \sigma^i T^a \psi 
	\; {\cal P}_{\chi_0 } \;
	\psi^\dagger \sigma^i T^a \chi \rangle
&\approx& 4 m_c \; {\cal O}^{ \chi_0 }_8({}^3S_1) \,,
\\ 
\langle \chi^\dagger (-\mbox{$\frac{i}{2}$} \tensor{\bf D} 
		\cdot \mbox{\boldmath $\sigma$}) \psi
	\; {\cal P}_{\chi_0 } \;
	\psi^\dagger (-\mbox{$\frac{i}{2}$} \tensor{\bf D} 
		\cdot \mbox{\boldmath $\sigma$}) \chi \rangle
&\approx& 12 m_c \; {\cal O}^{ \chi_0 }_1({}^3P_{0}) \,.
\end{eqnarray}

We now compare our results for the gluon fragmentation functions 
of the $\chi_{cJ}$ states with previous calculations.
 These fragmentation functions were first calculated 
by Braaten and Yuan\cite{Braaten-Yuan:Pwave}. 
They regularized the infrared divergence from the process 
$g^* \to c \bar c g$ by imposing a cut-off $|{\bf k}| > \Lambda$ 
on the momentum of the real gluon in the final state, and  
they used the covariant projection method to isolate the 
contributions to the fragmentation functions for $\chi_{c0}$,  
$\chi_{c1}$, and $\chi_{c2}$. They also
assumed implicitly that the renormalization of the NRQCD matrix element 
$\langle \chi^\dagger \sigma^i T^a \psi 
	{\cal P}_{c \bar c',c \bar c}
	\psi^\dagger \sigma^i T^a \chi \rangle$
was carried out in such a way that the net effect 
of terms on the NRQCD side of the matching condition
was simply to transform the infrared cutoff $\Lambda$ 
into an ultraviolet cutoff on NRQCD.  
The fragmentation functions 
 calculated in Ref. \cite{Braaten-Yuan:Pwave} 
have been checked by Cho, Wise, and Trivedi \cite{C-W-T},
who extended the calculation to the polarized
fragmentation functions of $g \to \chi_J$.
Our results differ from those 
 in Ref. \cite{Braaten-Yuan:Pwave} only in the coefficients of 
the $\delta (1-z)$ terms
in (\ref{D1-J}).  Braaten and Yuan obtained
 $Q_J' - (2J+1) \ln(\Lambda/m_c)$,
 where
$Q_0' = {13 \over 12}$, $Q_1' = {23 \over 8}$, and $Q_2' = {121 \over 24}$.
The numbers $Q_J$ and $Q_J'$ differ by ${5 \over 6}(2 J + 1)$.
Since the difference between the coefficients of
$\delta (1-z)$ is proportional to $2J+1$, it can be absorbed
into a shift in the value 
of  the color-octet matrix element
$\langle \chi^\dagger \sigma^i T^a \psi {\cal P}_{\chi_0 }
	\psi^\dagger \sigma^i T^a \chi \rangle.$
The relation between the matrix elements in the two calculations is 
\begin{eqnarray}
\langle \chi^\dagger \sigma^i T^a \psi 
	\; {\cal P}_{\chi_0 } \;
	\psi^\dagger \sigma^i T^a \chi \rangle^{(\mu)} \Bigg|_{\rm dim.reg.}
&\; = \;&
\langle \chi^\dagger \sigma^i T^a \psi 
	\; {\cal P}_{\chi_0 } \;
	\psi^\dagger \sigma^i T^a \chi \rangle^{(\Lambda)} \Bigg|_{\rm cutoff}
\nonumber \\
&& \hspace{-1in}
\;+\; {16 \alpha_s \over 81 \pi m_c^2}
\left( \ln {\mu \over 2 \Lambda}  + {5 \over 6} \right) 
\langle  \chi^\dagger (-\mbox{$\frac{i}{2}$} \tensor{\bf D} 
		\cdot \mbox{\boldmath $\sigma$}) \psi
	\; {\cal P}_{\chi_0 } \;
	\psi^\dagger (-\mbox{$\frac{i}{2}$} \tensor{\bf D} 
		\cdot \mbox{\boldmath $\sigma$}) \chi \rangle  \,.
\end{eqnarray}
Thus our calculation is consistent with that of Braaten and Yuan.
The differences in the short-distance coefficients are due simply 
to different definitions of 
the color-octet 
matrix element. 

The gluon fragmentation functions for $\chi_{cJ}$ have also been 
calculated by Ma \cite{Ma-Pwave}.
Ma used dimensional regularization to cut off the  infrared divergence 
from the process $g^* \to c \bar c g$.  He also used the 
covariant projection method to isolate the 
contributions to the fragmentation functions for $\chi_{c0}$,  
$\chi_{c1}$, and $\chi_{c2}$. 
His results differ from ours only in 
the numbers $Q_J$ multiplying the $\delta (1-z)$ terms in (\ref{D1-J}). 
He obtained $Q_0 = {1 \over 4}$, $Q_1 = {15 \over 8}$, 
and $Q_2  = {19 \over 8}$.
The differences between his values of $Q_J$ and ours are not proportional to
$2J +1$.  Therefore, 
they cannot be absorbed into a redefinition of the 
color-octet matrix element. 
We attribute the discrepancy to an inconsistency in Ma's calculation.
Dimensional regularization of the infrared divergence
involves analytically continuing integrals to 
$3 - 2 \epsilon$ dimensions,
while the standard covariant projection method 
uses projection matrices that are specific to 3 dimensions.

\section{Conclusions}

Dimensional regularization is the most convenient method for regularizing 
the infrared and ultraviolet divergences that arise in calculations
of quarkonium production and annihilation rates beyond leading order. 
The standard covariant projection method is incompatible with
dimensional regularization. An alternative method for calculating 
the short-distance coefficients in the NRQCD factorization formulas is
the threshold expansion method developed in Ref. \cite{Braaten-Chen}. 
In this paper, we have generalized this method to $N$ spatial dimensions,
so that dimensional regularization can be used consistently 
to regularize the infrared and ultraviolet divergences that arise in 
perturbative calculations.
We illustrated the method by calculating the 
color-octet terms of order $\alpha_s$ and the color-singlet terms 
of order $\alpha_s^2$ in the 
gluon fragmentation functions for arbitrary quarkonium states.  
We resolved a
discrepancy between two  previous calculations of the 
gluon fragmentation functions for the spin-triplet P-wave quarkonium states.  
This general and systematic method should be very useful in 
extending calculations of the production and decay rates of heavy quarkonium
states to higher orders in perturbation theory.

\acknowledgements

This work was supported in part by the U.S.
Department of Energy, Division of High Energy Physics, under 
Grant DE-FG02-91-ER40684.

%%%%%%%%%%%%%%%%%%%%%%%%%%%%%%  APPENDICES  %%%%%%%%%%%%%%%%%%%%%%%%%%%%%%

\appendix

\section{Nonrelativistic expansion of spinors}

This appendix is identical to  Appendix A of Ref. \cite{Braaten-Chen},
except that it includes only those formulas that generalize to 
$N$ spacial dimensions.
Formulas involving  Levi-Civita tensors that cannot be easily generalized 
are omitted.
We give the nonrelativistic expansions for 
the spinors of a heavy quark $c$ and antiquark $\bar c$ for arbitrary 
momentum $P=(P_0,P_1, \ldots, P_N)$.
We assume that the relative   momentum ${\bf q} = (q_1, \ldots, q_N)$
of the $c$ in the center-of-momentum (CM) frame  of the $c \bar c$ pair
is small compared to the quark mass $m_c$.  The momenta
$p$ and $\bar p$ of the $c$ and $\bar c$ can be written
\begin{mathletters}
\begin{eqnarray}
p \;=\; \mbox{$1 \over 2$} P \;+\; L {\bf q} \;,
\\
\bar p \;=\; \mbox{$1 \over 2$} P \;-\; L {\bf q} \;,
\end{eqnarray}
\end{mathletters}
where $P$ is the total momentum and $L$ is a Lorentz boost matrix.  From 
the mass-shell conditions, $p^2 = \bar p^2 = m_c^2$, we have
$P \cdot L{\bf q} = 0$ and $P^2 = 4 E_q^2$, where 
$E_q = \sqrt{m_c^2 + {\bf q}^2}$. 
The components of the momenta $P$ and $L {\bf q}$ in the CM frame 
of the pair are
\begin{mathletters}
\begin{eqnarray}
P^\mu \bigg|_{\rm CM} &\;=\;& ( 2 E_q , \; {\bf 0} ) \;,
\\
(L {\bf q})^\mu \bigg|_{\rm CM} &\;=\;& ( 0, \; {\bf q} ) \;.
\end{eqnarray}
\end{mathletters}
When boosted to an arbitrary frame in which
the pair has total  spacial momentum ${\bf P}$, these momenta are
\begin{mathletters}
\begin{eqnarray}
P^\mu &\;=\;&
\left( \sqrt{ 4 E_q^2 + {\bf P}^2} , \; {\bf P} \right) \;,
\\
(L {\bf q})^\mu &\;=\;& L^\mu_j \; q^j .
\end{eqnarray}
\end{mathletters}
The boost matrix $L^\mu_{\ j}$, which has one Lorentz index and one Cartesian
index, has components
\begin{mathletters}
\label{L-boost}
\begin{eqnarray}
L^0_{\ j} &\;=\;& {1 \over 2 E_q} P^j \; , 
\\
L^i_{\ j} &\;=\;& \delta^{ij} - {P^i P^j \over {\bf P}^2}
	\;+\; {P^0 \over 2 E_q} {P^i P^j \over {\bf P}^2}  \;.
\end{eqnarray}
\end{mathletters}
The contraction of the boost tensor $L^\mu_{\ i}$ with the Lorentz 
vector $P$ vanishes: $P_\mu L^\mu_{\ j} = 0$.
The contractions of two boost matrices in their Lorentz indices or in their
Cartesian indices have simple forms:
\begin{mathletters}
\begin{eqnarray}
g_{\mu \nu} L^\mu_{\ i} L^\nu_{\ j} &\;=\;& - \delta^{ij} ,
\label{id:gL}
\\
L^\mu_{\ i} L^\nu_{\ i} &\;=\;& - g^{\mu \nu} \;+\; {P^\mu P^\nu \over P^2} .
\label{id:LL}
\end{eqnarray}
\end{mathletters}

The representation for gamma matrices that is most convenient 
for carrying out the nonrelativistic expansion of a spinor 
is the Dirac representation:
\begin{equation}
\gamma^0  \;=\; 
\left( \begin{array}{cc} 
	1 &  0  \\ 
	0 & -1 
	\end{array} \right) , 
\qquad
\gamma^i \;=\;  
\left( \begin{array}{cc} 
	     0    & \sigma^i \\ 
	-\sigma^i &    0        
	\end{array} \right) .
\end{equation}
In the CM frame of the $c \bar c$ pair, the spinors for the $c$ and 
the $\bar c$ are
\begin{mathletters}
\label{spin-rest}
\begin{eqnarray}
u(p) \Bigg|_{\rm CM} 
&\;=\;& {1 \over \sqrt{E_q + m_c}}
\left( \begin{array}{c} 
	(E_q + m_c) \; \xi \\ 
	{\bf q} \cdot \mbox{\boldmath $\sigma$} \; \xi 
	\end{array} \right) ,
\label{u-rest}
\\
v(\bar p) \Bigg|_{\rm CM}
&\;=\;& {1 \over \sqrt{E_q + m_c}}
\left( \begin{array}{c} 
	- {\bf q} \cdot \mbox{\boldmath $\sigma$} \; \eta \\
	(E_q + m_c) \; \eta
	\end{array} \right) .
\label{v-rest}
\end{eqnarray}
\end{mathletters}
Color and spin quantum numbers on the Dirac spinors
and on the Pauli spinors $\xi$ and $\eta$ are suppressed.
When boosted to a frame in which the pair has total  spacial
momentum ${\bf P}$, the spinors for the $c$ and $\bar c$ are
\begin{mathletters}
\begin{eqnarray}
u(p) 
&\;=\;& {1 \over \sqrt{4 E_q (P_0 + 2 E_q) (E_q + m_c)}}
\left( 2 E_q + \rlap/{\! P} \gamma_0 \right)
\left( \begin{array}{c} 
	(E_q + m_c) \; \xi \\ 
	{\bf q} \cdot \mbox{\boldmath $\sigma$} \; \xi 
	\end{array} \right) ,
\label{u-spinor}
\\
v(\bar p) 
&\;=\;& {1 \over \sqrt{4 E_q (P_0 + 2 E_q) (E_q + m_c)}}
\left( 2 E_q + \rlap/{\! P} \gamma_0 \right)
\left( \begin{array}{c} 
	- {\bf q} \cdot \mbox{\boldmath $\sigma$} \; \eta \\
	(E_q + m_c) \; \eta
	\end{array} \right) .
\label{v-spinor}
\end{eqnarray}
\end{mathletters}
These spinors are normalized so that $\bar u u = - \bar v v = 2 m_c$ if the 
Pauli spinors are normalized so that $\xi^\dagger \xi = \eta^\dagger \eta = 1$.

The independent quantities that can be formed by sandwiching
3 or fewer Dirac matrices between $\bar u(p)$ and $v(\bar p)$ are
\begin{mathletters}
\label{bispinors}
\begin{eqnarray}
&& \bar u(p) v(\bar p) \;=\; 
- 2 \; \xi^\dagger ({\bf q} \cdot \mbox{\boldmath $\sigma$}) \eta ,
\\
&& \bar u(p) \gamma^\mu v(\bar p) \;=\; L^\mu_{\ j} \;
\left( 2 E_q \; \xi^\dagger \sigma^j \eta \;-\; {2 \over E_q + m_c} \; q^j \;
	\xi^\dagger ( {\bf q} \cdot \mbox{\boldmath $\sigma$} ) \eta \right) 
\;,                                                                              
\label{bispinor-gam}
\\
&& \bar u(p) ( \gamma^\mu \gamma^\nu
           -\gamma^\nu \gamma^\mu ) v(\bar p) \;=\; 
( P^\mu L^\nu_{\ j} - P^\nu L^\mu_{\ j} ) 
\left( {2 m_c \over E_q} \; \xi^\dagger \sigma^j \eta 
	\;+\; {2 \over E_q(E_q + m_c)} q^j \; 
	\xi^\dagger ({\bf q} \cdot \mbox{\boldmath $\sigma$}) \eta \right)
\nonumber \\
&& \hspace{2in}
\;+\;  L^\mu_{\ j} \; L^\nu_{\ k} \; 
	\xi^\dagger \{ [ \sigma^j, \sigma^k ], 
		{\bf q} \cdot \mbox{\boldmath $\sigma$} \} \eta,
\\
&& \bar u(p) ( \gamma^\mu \gamma^\nu \gamma^\lambda
           -\gamma^\lambda \gamma^\nu \gamma^\mu ) v(\bar p) 
\nonumber \\
&& \hspace{.5in} 
\;=\; L^\mu_{\ i} \; L^\nu_{\ j} \; L^\lambda_{\ k} 
\Bigg( - E_q \; \xi^\dagger \{ [ \sigma^i, \sigma^j ] , \sigma^k \} \eta 
\;+\; {q^i \over E_q + m_c} \; \xi^\dagger \{ [ \sigma^j, \sigma^k ] , 
			{\bf q} \cdot \mbox{\boldmath $\sigma$} \} \eta 
\nonumber \\ 
&& \hspace{1.5in}
\;+\; {q^j \over E_q + m_c} \; \xi^\dagger \{ [ \sigma^k, \sigma^i ] , 
			{\bf q} \cdot \mbox{\boldmath $\sigma$} \} \eta 
\;+\; {q^k \over E_q + m_c} \; \xi^\dagger \{ [ \sigma^i, \sigma^j ] , 
			{\bf q} \cdot \mbox{\boldmath $\sigma$} \} \eta 
\Bigg)
\nonumber \\
&& \hspace{1in}
\;-\; {2 \over E_q} 
\left( P^\mu L^\nu_{\ i} L^\lambda_{\ j}
	+ L^\mu_{\ i} L^\nu_{\ j} P^\lambda 
	+ L^\mu_{\ j} P^\nu L^\lambda_{\ i} \right)
\left( \xi^\dagger q^i \sigma^j \eta 
	\;-\; \xi^\dagger q^j \sigma^i \eta \right) .
\end{eqnarray}
\end{mathletters}
The simplest way to derive these formulas is to use the identities 
\begin{mathletters}
\begin{eqnarray}
\left( 2 E_q + \gamma_0 \rlap/{\! P} \right)
	\left( 2 E_q + \rlap/{\! P} \gamma_0 \right) 
&=& 4 E_q ( P_0 + 2 E_q ),
\\
\left( 2 E_q + \gamma_0 \rlap/{\! P} \right) \gamma^\mu
	\left( 2 E_q + \rlap/{\! P} \gamma_0 \right) 
&=& 4 E_q ( P_0 + 2 E_q )
\left( {P^\mu \over 2 E_q} \; \gamma_0 + L^\mu_{\ j} \; \gamma^j \right).
\end{eqnarray}
\end{mathletters}
Using the expressions for the spinor factors given in (\ref{bispinors}), 
it is easy to carry out the nonrelativistic
expansions in powers of ${\bf q}$.

%%%%%%%%%%%%%%%%%%%%%%%%%%%%%%  REFERENCES  %%%%%%%%%%%%%%%%%%%%%%%%%%%%%%

\vfill \eject
%%%%%%%%%%%%%%%%%%% FIGURE CAPTIONS %%%%%%%%%%%%%%%%%%%%%%%%%

\begin{figure}
{ Fig.~1. 
The lowest-order Feynman diagram in QCD for 
$g^* (l) \to c (p) + \bar{c} ( \bar{p} ) $. }
\end{figure}

\begin{figure}
{ Fig.~2. 
The lowest-order Feynman diagram for the NRQCD matrix element 
$\langle \chi^\dagger \sigma^k T^a \psi  
	{\cal P}_{c \bar c', c \bar c}
	\psi^\dagger \sigma^k T^a \chi \rangle$.
The lines entering at the left and leaving at the right	
come from a term of the form $| c \bar c' \rangle \langle c \bar c |$
in the projection operator ${\cal P}_{c \bar c', c \bar c}$.} 
\end{figure}

\begin{figure}
{Fig.~3. 
The lowest-order Feynman diagrams in QCD for 
$g^* (l) \to c (p) + \bar{c} ( \bar{p} ) + g (k)$
when the $c \bar c$ pair is in a color-singlet state. }
\end{figure}

\figure{Fig.~4.
The lowest-order Feynman diagrams for the NRQCD matrix element
$\langle \chi^\dagger \sigma^k T^a \psi  
	{\cal P}_{c \bar c', c \bar c}
	\psi^\dagger \sigma^k T^a \chi \rangle$
when the projection operator ${\cal P}_{c \bar c', c \bar c}$
requires the $c \bar c$ pairs to be in color-singlet states.
The lines entering at the left and leaving at the right	
come from a term of the form $| c \bar c' g \rangle \langle c \bar c g |$
in the projection operator. }
  
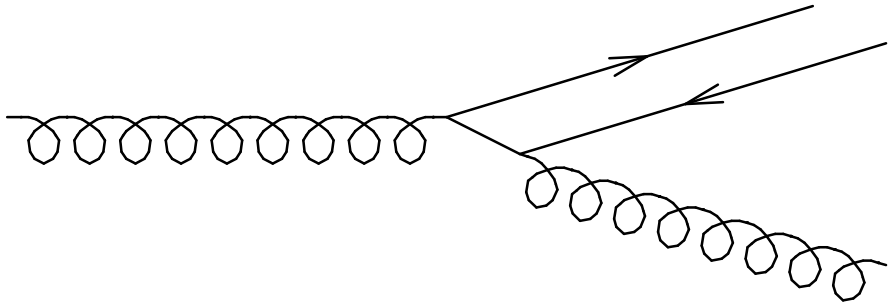
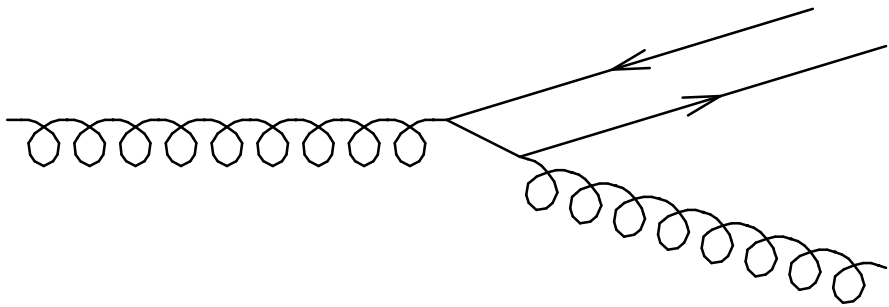
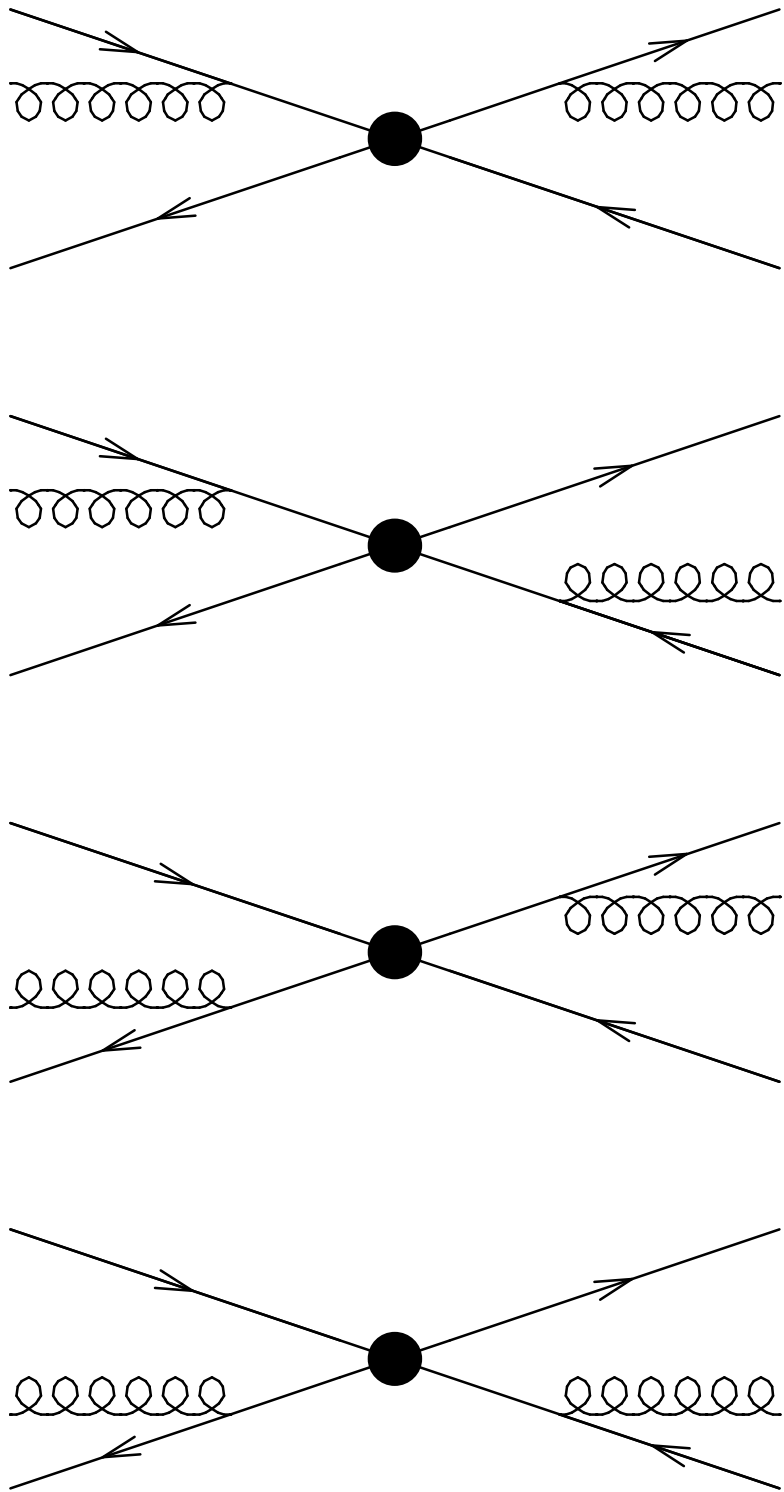
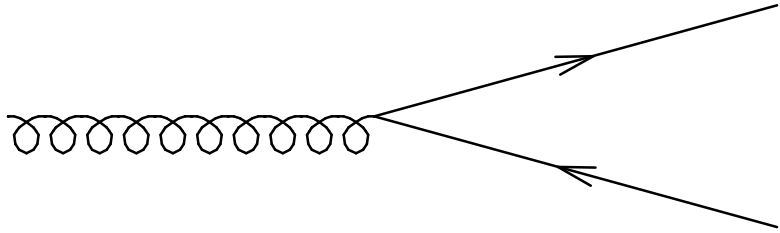
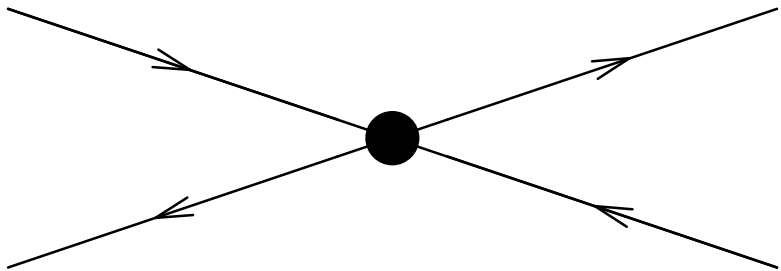
\end{document}